\documentclass[preprint]{imsart}

\RequirePackage{amsthm,amsmath,amsfonts,amssymb,booktabs}
\RequirePackage[authoryear]{natbib}
\RequirePackage[colorlinks,citecolor=blue,urlcolor=blue]{hyperref}
\RequirePackage{graphicx}
\RequirePackage{changepage}

\usepackage[usenames,dvipsnames]{xcolor}
\usepackage{amsmath, amsfonts, amssymb}

\newcommand{\bi}{\begin{itemize}}
\newcommand{\ei}{\end{itemize}}

\newcommand{\be}{\begin{enumerate}}
\newcommand{\ee}{\end{enumerate}}

\newcommand{\cutpt}{\gamma}
\newcommand{\cutpts}{\boldsymbol{\gamma}}
\newcommand{\CorM}{\boldsymbol{R}}
\newcommand{\CovM}{\boldsymbol{\Sigma}}

\newcommand{\loc}{\mathbf{s}}

\newcommand{\norm}{\mathrm{N}}

\newcommand{\Scal}{\mathcal{S}}

\newcommand{\0}{\mathbf{0}}

\newcommand{\A}{\mathbf{A}}
\newcommand{\D}{\mathbf{D}}

\newcommand{\abf}{\mathbf{a}}
\newcommand{\I}{\mathbf{I}}

\newcommand{\hh}{\mathbf{h}}
\newcommand{\HH}{\mathbf{H}}

\newcommand{\M}{\mathbf{M}}

\newcommand{\w}{\mathbf{w}}

\newcommand{\s}{\mathbf{s}}

\newcommand{\X}{\mathbf{X}}
\newcommand{\x}{\mathbf{x}}
\newcommand{\Y}{\mathbf{Y}}

\newcommand{\Z}{\mathbf{Z}}
\newcommand{\Zrm}{\mathrm{Z}}

\newcommand{\betabf}{\boldsymbol{\beta}}
\newcommand{\Gammabf}{\boldsymbol{\Gamma}}
\newcommand{\gammabf}{\boldsymbol{\gamma}}

\newcommand{\deltabf}{\boldsymbol{\delta}}
\newcommand{\phibf}{\boldsymbol{\phi}}

\newcommand{\Lambdabf}{\boldsymbol{\Lambda}}
\newcommand{\mubf}{\boldsymbol{\mu}}
\newcommand{\nubf}{\boldsymbol{\nu}}
\newcommand{\thetabf}{\boldsymbol{\theta}}

\newcommand{\etabf}{\boldsymbol{\eta}}
\newcommand{\epsilonbf}{\boldsymbol{\epsilon}}
 
\newcommand{\sigmabf}{\boldsymbol{\sigma}}

\startlocaldefs
\endlocaldefs

\begin{document}

\begin{frontmatter}
\title{Land-Use Filtering for Nonstationary Spatial Prediction of Collective
	Efficacy in an Urban Environment}
%\title{A sample article title with some additional note\thanksref{T1}}
\runtitle{Land-Use Filtering}
%\thankstext{T1}{A sample of additional note to the title.}

\begin{aug}
%%%%%%%%%%%%%%%%%%%%%%%%%%%%%%%%%%%%%%%%%%%%%%%
%% Only one address is permitted per author. %%
%% Only division, organization and e-mail is %%
%% included in the address.                  %%
%% Additional information can be included in %%
\author[A]{\fnms{J. Brandon} \snm{Carter}\ead[label=e1]{carterjb@utexas.edu}\orcid{0000-0003-1687-0564}},
\author[B]{\fnms{Christopher R.} \snm{Browning}\ead[label=e4]{browning.90@osu.edu}\orcid{0000-0001-7807-3849}},
 \author[C]{\fnms{Bethany} \snm{Boettner}\ead[label=e5]{boettner.6@osu.edu}\orcid{0000-0002-4390-9447}},
\author[B]{\fnms{Nicolo} \snm{Pinchak}\ead[label=e3]{pinchak.5@osu.edu}\orcid{0000-0003-1853-1558}},
\and
\author[A]{\fnms{Catherine A.} \snm{Calder}\ead[label=e2]{calder@austin.utexas.edu}\orcid{0000-0002-4459-1418}}
% %%%%%%%%%%%%%%%%%%%%%%%%%%%%%%%%%%%%%%%%%%%%%%
% %% Addresses                                %%
% %%%%%%%%%%%%%%%%%%%%%%%%%%%%%%%%%%%%%%%%%%%%%%
 \address[A]{Department of Statistics and Data Sciences, University of Texas at Austin\printead[presep={,\ }]{e1,e2}}

 \address[B]{Department of Sociology, The Ohio State University\printead[presep={,\ }]{e4,e3}}

 \address[C]{Population Research Institute, The Ohio State University\printead[presep={,\ }]{e5}}
\end{aug}

\begin{abstract}
%
%Over the last decade, nonstationary spatial modeling strategies that leverage spatially-referenced covariates for parsimonious model specification have
%been developed for applications in the environmental and climate sciences.
	Collective efficacy -- the capacity of communities to exert social control toward the realization of their shared goals -- is a foundational concept in the urban sociology and neighborhood effects literature.  Traditionally, empirical studies of collective efficacy use large sample surveys to estimate collective efficacy of different neighborhoods within an urban setting.  Such studies have demonstrated an association between collective efficacy and local variation in community violence, educational achievement, and health.  Unlike traditional collective efficacy measurement strategies, the Adolescent Health and Development in Context (AHDC) Study implemented a new approach, obtaining spatially-referenced, place-based ratings of collective efficacy from a representative sample of individuals residing in Columbus, OH. In this paper, we introduce a novel nonstationary spatial model for interpolation of the AHDC collective efficacy ratings across the study area which leverages administrative data on land use.  Our constructive model specification strategy involves dimension expansion of a latent spatial process and the use of a filter defined by the land-use partition of the study region to connect the latent multivariate spatial process to the observed ordinal ratings of collective efficacy. Careful consideration is given to the issues of parameter identifiability, computational efficiency of an MCMC algorithm for model fitting, and fine-scale spatial prediction of collective efficacy.  

\end{abstract}


\end{frontmatter}
%%%%%%%%%%%%%%%%%%%%%%%%%%%%%%%%%%%%%%%%%%%%%%
%% Please use \tableofcontents for articles %%
%% with 50 pages and more                   %%
%%%%%%%%%%%%%%%%%%%%%%%%%%%%%%%%%%%%%%%%%%%%%%
%\tableofcontents

%%%%%%%%%%%%%%%%%%%%%%%%%%%%%%%%%%%%%%%%%%%%%%
%%%% Main text entry area:
\section{Introduction}
\label{sec:intro}
For decades, social science research has investigated how neighborhood residents can mobilize to address local problems
\citep{shaw_mckay1942, jacobs1961}.  The theory of neighborhood collective efficacy has been highly influential in this respect \citep{sampson_etal1997}. Uniting \citeauthor{coleman1988}'s
\citetext{\citeyear{coleman1988}} concept of social capital and \citeauthor{bandura1982}'s \citetext{\citeyear{bandura1982, bandura1986}} research on personal and group-level self-efficacy, neighborhood collective
efficacy is a construct  capturing the collective capacity of community members to exert
social control toward the realization of their shared goals (e.g., low levels of crime). 

Empirical tests of the neighborhood collective efficacy theory often rely on data collected through sample surveys administered to residents of different neighborhoods within an urban center.   Collective efficacy survey instruments vary, but typically ask urban residents to report perceptions of local trust, monitoring, and intervention norms in their neighborhood. We refer to these three categories of questions -- "trust," "observation," and "defense," respectively -- as the \textit{components} of collective efficacy. 
%(i.e., the presence of ``eyes on the street'' \citep{jacobs1961})
Reports on these components are then aggregated within geographic units consistent with the notion of neighborhoods (e.g., census tracts).  For
example, the Project on Human Development in Chicago Neighborhoods (PHDCN), conducted in the mid 1990s, used
neighborhood ratings from nearly 8,000 Chicago residents clustered within 343
census-based neighborhoods to capture collective efficacy
\citetext{\citealp{raudenbush_sampson1999}; see also
\citealp{matsueda_drakulich2016, wickes_etal2019}}.  In the PHDCN and other studies, multilevel/hierarchical regression models were used to capture between-neighborhood variation in collective efficacy through random effects, which are assumed to be spatially independent or dependent over the study region depending on the study design.  
%Survey instruments in empirical collective efficacy research vary, but generally include one or more questions about trust among neighbors, whether %neighbors come to the defense of each other, and the capacity for social control (i.e., the presence of ``eyes on the street'' \citep{jacobs1961}).
Studies of neighborhood-level collective efficacy have found inverse associations with a host of neighborhood problems such as rates of
homicide \citep{sampson_etal1997}, intimate partner violence
\citep{browning2002}, child maltreatment \citep{molnar_etal2015}, and chronic
disease \citep{cohen_etal2006}, even after adjusting for neighborhood
sociodemographic factors.

In this paper, we extend the notion of collective efficacy as a continuously-indexed
spatial feature of an urban area, which is defined by the collective impressions of
places individuals visit as part of their normal, everyday routine.  Using
point-referenced collective efficacy data collected as part of the Adolescent Health and
Development in Context (AHDC) Study, described in detail in Section \ref{sec:eda}, we
estimate the components of collective efficacy across the city of Columbus, OH.  Our 
approach also uses taxation-based records on the land use of parcels -- small geographic 
units often used as a proxy for a ``place.''  Consistent with recent work suggesting that 
land-use compositions affect the levels of collective efficacy within neighborhoods 
\citep{corcoran_etal2018}, we use the category of land use to model spatial variation in
the mean of the collective efficacy component processes.  Our novel contribution from a 
statistical perspective is that we capture land-use driven  (i.e. driven by spatially-referenced
covariate information), second-order nonstationarity through a dimension-expansion strategy,
%that leverages spatially-referenced covariate information,
building upon existing approaches for covariate-driven nonstationary spatial modeling  
\citep{calder2008, schmidt_etal2011, reich_etal2011, vianna_etal2014, ingebrigtsen2014, risser_calder2015, risser_etal2019}.
In a similar spirit to the dimension expansion approach to nonstationary spatial modeling 
of \citet{bornn_etal2012}, we augment the dimension of the latent collective efficacy 
component processes to the number of land-use categories and model a multivariate, stationary spatial process as if spatially-referenced data on land-use specific collective efficacy components were observable everywhere.
We then relate the observations to the higher dimensional random process using a  filter defined by the land-use partitioned study area. 
We show that our land-use filtering model provides a better fit to the AHDC collective efficacy data than a traditional spatial generalized linear mixed model with a single latent second-order stationary Gaussian process \citep{banerjee_etal2014}.   

Our proposed methodology advances the measurement of collective efficacy in two ways.  First, it is compatible with the more cost effective data collection strategy employed in the AHDC Study, where study participants report on collective efficacy levels at their routine activity locations, of which most have between five and eight. In addition, our model readily allows point-level prediction of the components of collective efficacy across the
entire study area to better understand within neighborhood variation in collective efficacy. Point-level variability in collective efficacy may help explain phenomena such as crime which has been shown to concentrate at particular locations within neighborhoods \citep{weisburd_etal2016}.  Finally, we note that while developed for the study of collective efficacy, our
methodology can be readily applied in other spatial prediction settings where the study area can be partitioned into an arbitrary number of land-use categories.

The outline of the paper is as follows. In Section~\ref{sec:eda}, we introduce the AHDC Study and the collective efficacy data.  In addition, we describe exploratory analyses of the data that motivate our novel land-use filtering methodology.  Section~\ref{sec:methods} introduces our land-use filtering model and details land-use filtering as
applied to our ordinal response variable. We provide a summary of inferences and predictions
from our fitted model to the AHDC data in Section~\ref{sec:results} and explore model performance when data are generated from a land-use filtering process through a simulation study in Section~\ref{sec:sim_study}. Lastly, we
discuss implications of our modeling and data collection choices in
Section~\ref{sec:discuss}.

\section{Data and exploratory analyses}
\label{sec:eda}
%
%The data utilized are survey responses from caregivers of adolescents enrolled
%in the Adolescent Health and Development in Context (AHDC) study. 

In this section, we describe the AHDC data in more detail and visually summarize the spatial patterning of the ratings of the components of collective efficacy.  We also introduce the land-use data, which drive our proposed spatial filtering strategy for smoothing the ratings across the study area.  This section concludes with summaries of preliminary models that motivate the more complex modeling strategy introduced in Section~\ref{sec:methods}. 

\subsection{AHDC data}
\label{sec:ahdc_data}
The AHDC Study is a longitudinal data collection project designed to improve understanding of how social, psychological, and biological processes shape youth developmental outcomes.  Participants in the study are members of a representative sample of households with youth aged 11 to 17 residing within the I-270 belt loop in Franklin County, OH, which contains the city of Columbus and some of its interior suburbs.  In this paper, we use data from AHDC Wave 1, which was collected between 2014-2016.  For each sampled household, one randomly-sampled youth aged 11-17 was selected for participation in the study. Upon enrollment, the primary caregiver, typically, but not always, the mother of the enrolled youth, filled out an
entrance questionnaire which included questions on family and household
composition, alcohol and substance use, employment and income, health, and
social support.  In addition to answering these demographic and socioeconomic
background questions, caregivers also listed locations they frequent as part of their
everyday routine. Caregivers indicated the location type (e.g. home, kid's
school, friends' and relatives' houses, work, grocery stores, etc.) and when
they typically spend time at the location (e.g.  daytime, nighttime, weekdays,
weekends).  They also answered multiple questions on each reported location to assess the social climate. These location ratings are the key data used in this current analysis to understand
spatial variation in collective efficacy. From the location assessments, we focus on three collective efficacy questions posed as statements to which the caregiver rated agreement on a five-point
Likert scale: \textit{If someone was being threatened near [location], other people around would come to their defense. You can trust people on the streets in the area near [location]. There are usually people watching what's happening in the area near [location].}\footnote{The first two statements had response options "strongly agree," "agree," "neither agree nor disagree," "disagree," and "strongly disagree," while the last statement had responses "never," "almost never," "sometimes," "fairly often" and "very often." Each response was displayed in the order listed above next to the corresponding numeric value 1-5. In fitting the model we recoded the responses so that 1 corresponded to the least affirmative value (never or strongly disagree) and 5 the most affirmative value (very often and strongly agree) for all three components.} We refer to the responses to these questions and their corresponding latent spatial processes, respectively, as "defense," "trust," and "observation." While
location ratings outside of the study area were provided, we restrict our analysis to the 4526 locations within the I-270 belt loop. 
Out of the 1369 caregivers in the data set, the number of reports by a single caregiver ranged from 1 to 27, with an average of 6.8 reports. With a few exceptions\footnote{Due to a technical error in the administration of the survey, the question about trust was omitted from the initial surveys resulting in fewer ratings for the trust component as compared to defense and observation.}, all reports included ratings on all three components of collective efficacy along with the time of day the location is visited (daytime, nighttime, or both) and the day of week (weekday, weekend, or both). Caregivers were asked to rate their home location for both daytime and nighttime separately.

% The goal of the analysis is to estimate the latent level of each component of
% collective efficacy from the point referenced survey responses across the
% metropolitan area of Columbus, Ohio within the 270 belt loop. Previous studies
% of collective efficacy have primarily focused on surveying participants on
% their own neighborhood. Often the level of collective efficacy is estimated for
% areal units, either at the block group or census tract level
% \citep{sampson_etal1997, raudenbush_sampson1999, savitz_raudenbush2009,
% matsueda_drakulich2016, wickes_etal2019}.  By asking individuals to rate
% additional routine locations beyond their own neighborhood we seek to gain
% greater insight into the variation of collective efficacy within the block
% group or census tract level.  We expect aggregate summaries of our model to be
% similar to traditional models, however, our survey and statistical methods give
% the advantage of understanding collective efficacy at a finer resolution within
% the block group and census tract. 
%
\begin{figure}[ht!]
\centering
\begin{minipage}{.5\textwidth}
	\centering
	\includegraphics[width=.85\linewidth]{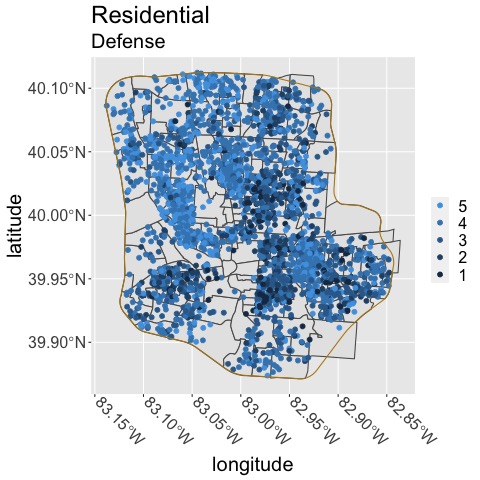}
\end{minipage}%
\begin{minipage}{.5\textwidth}
	\centering
	\includegraphics[width=.85\linewidth]{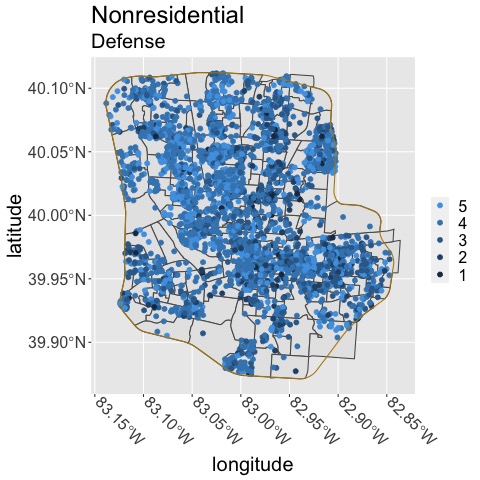}
\end{minipage}
\begin{minipage}{.5\textwidth}
	\centering
	\includegraphics[width=.85\linewidth]{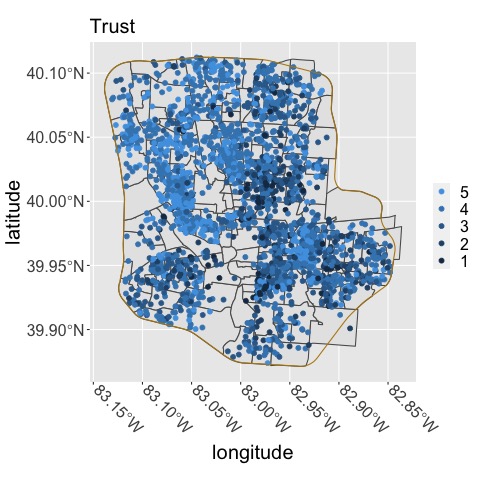}
\end{minipage}%
\begin{minipage}{.5\textwidth}
	\centering
	\includegraphics[width=.85\linewidth]{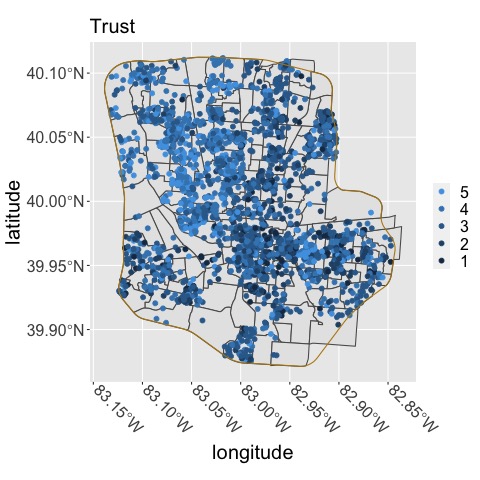}
\end{minipage}
\begin{minipage}{.5\textwidth}
	\centering
	\includegraphics[width=.85\linewidth]{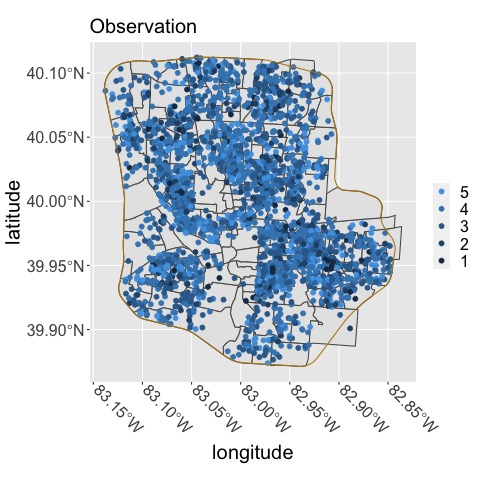}
\end{minipage}%
\begin{minipage}{.5\textwidth}
	\centering
	\includegraphics[width=.85\linewidth]{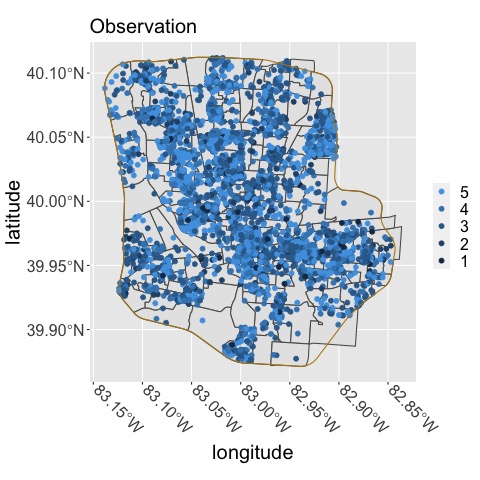}
\end{minipage}
\caption{Survey responses for the defense (top row), trust (middle row), and observation (bottom row) components plotted for residential
	locations (left column) and nonresidential locations (right column). The gray lines show census tract
	boundaries and the orange line is the 270 belt loop. Rated locations have been
	jittered within census block group to preserve the anonymity of study participants. Due to the jittering, some locations are plotted outside the I-270 belt loop.  The plots were created in \texttt{R} \citep{r2022, r_sf2018, r_tidyverse2019}.}
\label{fig:compare_loctype}
\end{figure}
Figure \ref{fig:compare_loctype} shows the locations to which the ratings of defense, trust, and observation correspond, plotted separately for residential and nonresidential locations (as derived from the caregiver's report on the location type). For the defense responses, the residential ratings are consistent with expectations if a spatially-dependent process underlies the ordinal ratings.  Spatial clustering of higher defense ratings in residential areas are apparent in the northwest quadrant and a pocket of residences relative to the lower
ratings of defense elsewhere in the city.  Nonresidential locations also appear to exhibit
the same general spatial patterning, yet not as well visually pronounced. While it is not always appropriate to draw conclusions about the strength of spatial dependence in a spatially-structured process underlying spatially-resolved oridinal data (see Sections~\ref{sec:motivation} and~\ref{sec:ident} for a discussion of this issue), the visual differences are indicative of the process underlying the ratings.  

A similar pattern is evident in Figure~\ref{fig:compare_loctype} for the trust and observation components -- the spatial patterning of ratings for residential and nonresidential locations are distinct. For trust (middle row), both residential and nonresidential ratings exhibit clear spatial patterning, but the pattern is not identical across both location types. In contrast, the observation ratings (bottom row) visually show less of a distinctive spatial pattern. It is difficult from visual inspection of these maps alone to draw conclusions about the underlying spatial process that give rise to the observed ratings across residential and nonresidential locations. Yet, for all three components, it appears that allowing the mean process to vary as a parametric function of observed, spatially-referenced covariates would not capture the differences in the spatial patterning of observed ratings across residential and nonresidential locations.

\subsection{Land-use data}
\label{sec:land-use_data}
\begin{figure}[ht]
\centering
	\includegraphics[width=.8\linewidth]{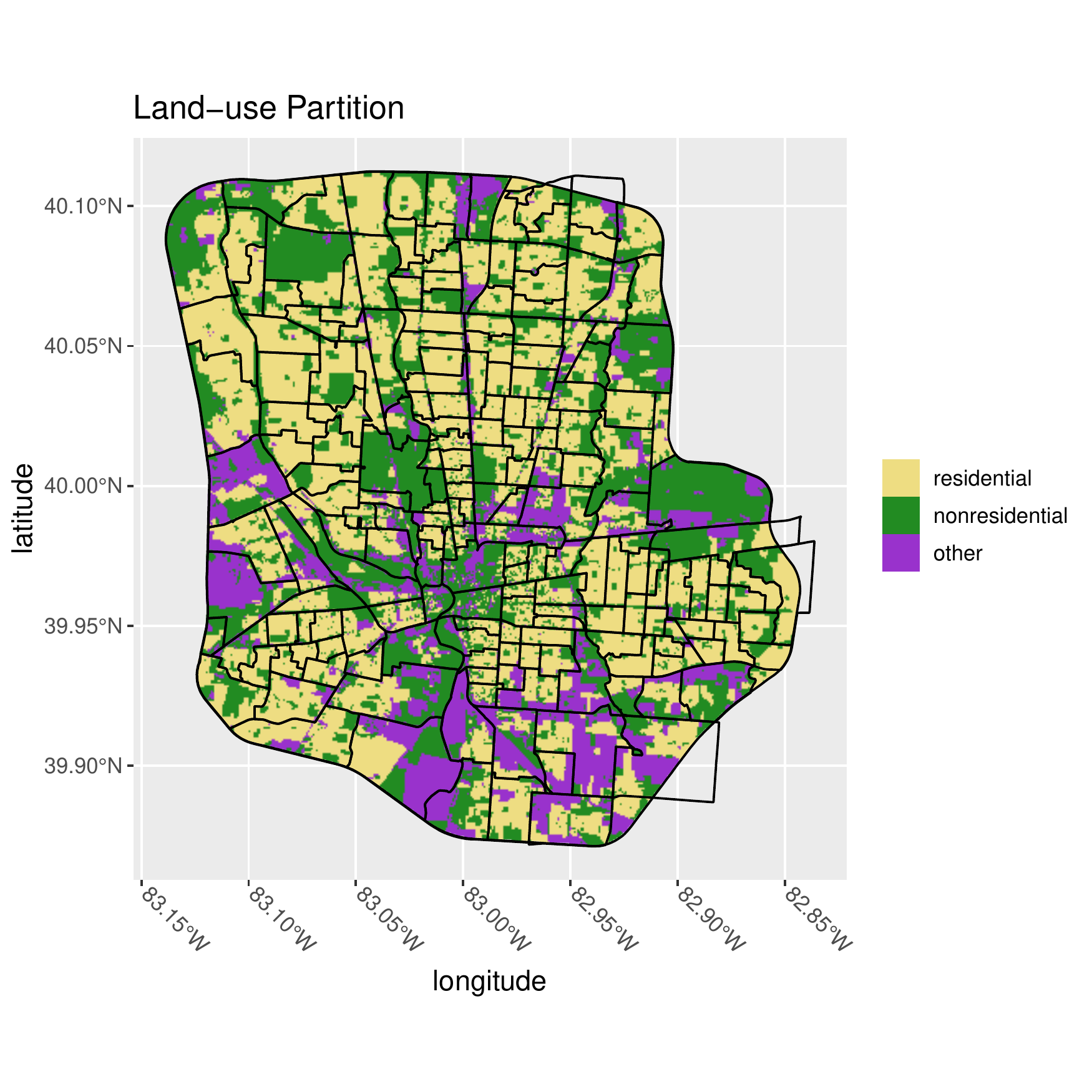}
\caption{Partition of the study region into land-use categories. Each cell color is determined by the tax code of the nearest parcel}
\label{fig:landuse_partition}
\end{figure}

In order to account for the effects of land use on differences in the spatial dependence structure of the components of collective efficacy across the study region when predicting collective efficacy components at unrated locations,
% , we 
% different spatial processes across land use types, we propose to expand our univariate
% response to a multivariate outcome, where the spatial process for each land-use
% category is only observed at locations with the corresponding land-use type. We
% call this form of dimension expansion and partial observation of spatial
% processes masked by land category - land-use filtering.
%
% We seek to use our land-use filtering model to predict the 
% collective efficacy component processes at unobserved locations across the entire
% study area. In order to do so, 
we require the land-use designation of all prediction locations, which is not available in the AHDC data.   In particular, caregivers' reported location types are straightforward to classify as residential and
nonresidential, however, we need an objective and systematic method to
categorize all other locations in the study area not reported on. To this end, we utilized the May 2014
parcel data for the study area, downloaded from the Franklin County Auditor's publicly accessible FTP site \citep{parcel2014}. Each parcel has a designated tax code which we
classified into one of three categories: residential, nonresidential, and other. We
introduce the ``other'' category to distinguish nonresidential locations with
minimal social presence, such as warehouses, quarries, parking lots and
industrial centers, from locations where a social presence is expected. 
Figure~\ref{fig:landuse_partition} shows the partition of the study area into residential,
nonresidential, and other categories as informed by the taxcodes of the parcel data.
Most
locations in the AHDC data lie within the boundaries of a parcel and received the same
classification as the parcel within which they lie. The coordinates of other
AHDC-reported locations fall on the road network and do not lie within the boundaries of a
parcel. For these locations we assigned the land-use category of the nearest parcel.
After initial assignments were made, we performed a second sweep of the data to verify
that the reported caregiver location type matched the assigned parcel land-use
type. Incongruencies existed between reported location types and assigned
land-use types (e.g. location reported by caregiver as ``neighborhood,'' but
assigned to parcel with tax designation ``warehouse'' and land-use category
``other''). For such locations, we reassigned the land-use category if any of
the next 5 nearest parcels had a tax code and land-use assignment that was
coherent with the reported location type. We deemed locations without a nearby
congruent parcel as a reporting error and did not include them in the analysis. Table~\ref{tab:ratings_counts} gives the total number of ratings and locations used in our analysis broken down by collective efficacy component and land-use category.

\begin{table}[ht]
\caption{Breakdown of the number of locations, $m$, and ratings, $n$, by collective efficacy component. The subscripts on $m$ and $n$ further breakdown location and ratings counts by land-use type, with 1 for residential, 2 for nonresidential and 3 for ``other''.}
\label{tab:ratings_counts}
\centering
\begin{tabular}{lrrrrr}
  \toprule
  & Defense & Trust & Observation\\
\midrule
  $m$ & 3867 & 3040 & 3826 \\
  \midrule
    $m_1$ & 1770 & 1555 & 1777\\
  $m_2$ & 2045 & 1447 & 1997\\
  $m_3$ & 52 & 38 & 52\\
 \midrule
 $n$ & 7580 & 5865 & 7548 \\
 \midrule
  $n_1$ & 2906 & 2653 & 2917\\
  $n_2$ & 4616 & 3170 & 4573\\
  $n_3$ & 58 & 42 & 58 \\
   \bottomrule
\end{tabular}
\end{table}

\subsection{Preliminary exploratory modeling}
\label{sec:prelim_model}
To better understand the difference in spatial variation in collective efficacy components by land-use categories, we fit separate univariate  Bayesian spatial ordinal regression
models to each component of collective efficacy by land-use category (excluding the ``other'' category which has only approximately 50 ratings for each component). That is, we fit six models corresponding to the three components of collective efficacy (defense, trust, and observation) by the two land-use categories (residential and nonresidential, as determined by the parcel data).  Section~\ref{sec:methods} provides details on the specification of 
spatial ordinal regression models but here we focus on inferences on key parameters describing the nature of the spatial dependence as a justification of our proposed model. Because Bayesian spatial ordinal regression models are partially identifiable (discussed in Section~\ref{sec:ident}), we clarify that inferences on the spatial parameters cannot be interpreted as characterizing the true latent continuously-indexed spatial process; rather, these inferences serve as evidence of distinct spatial patterns by land-use category. With this caveat in mind,
let $\thetabf=(\phi, \tau^2)$ be the vector of spatial dependence parameters characterizing the exponential covariance function of the latent spatial process, where $\phi$ represents the spatial range and $\tau^2$ is the parameter that governs the proportion of variance due to spatial variation, as opposed to independent error (in the Bayesian ordinal regression model the variance of the independent error is fixed at one).  
% \todo{I don't think this is needed here:  The correlation matrix for all six models is of
% the form $\CorM^*(\thetabf)=(1-\kappa)\CorM(\phi) + \kappa\I$, where $\kappa
% = \frac{1}{1+\tau^2}$, and $\I$ is the identity matrix. Here, $\CorM(\phi)$ is a correlation matrix defined by exponential correlation
% function $\rho(\loc,\loc'; \phi) = \exp(-\phi * d_{\loc,\loc'})$, where
% $d_{\loc,\loc'}$ is the Euclidean distance between observations at locations
% $\loc$ and $\loc'$. We placed $\mathrm{Cauchy}^+(0,1)$ priors on $\tau^2$ and
% $\phi$, where $\mathrm{Cauchy}^+$ denotes the the Cauchy distribution truncated
% to the support $(0,\infty)$.}
For all models, we adjust for the time-of-day
(daytime, nighttime, or both) and the day-of-week (weekday only, weekend only,
or mixed) variables in the mean function as indicated by the caregiver in the
survey response. 
% \todo{Again:  I think this isn't needed.  A flat prior was placed on the cut points and a $\norm(0,100)$
% prior on each $\beta$ in the mean function.} 
We obtained draws from the posterior distribution using the MCMC scheme described in Section~\ref{sec:mcmc}.

% latex table generated in R 4.1.1 by xtable 1.8-4 package
% Tue Oct  5 13:00:00 2021
\begin{table}[ht]
\caption{Estimates of the marginal posterior mean of the covariance parameters from Bayesian spatial ordinal models for the defense, trust, and observation components fitted separately to the residential and
	nonresidential ratings.}
\label{tab:prelim_model}
\centering
\begin{tabular}{rrrrrrrrr}
  \toprule
  &\multicolumn{2}{c}{Defense} & &\multicolumn{2}{c}{Trust} & & \multicolumn{2}{c}{Observation}\\
  \cmidrule{2-3} \cmidrule{5-6} \cmidrule{8-9}
 & $\phi$ & $\tau^2$ & & $\phi$ & $\tau^2$ & & $\phi$ & $\tau^2$\\ 
  \midrule
Residential & 334.29 & 2.08 &  & 443.68 & 3.03 & & 2156.17 & 2.06 \\ 
  Nonresidential & 53.28 & 0.15 &  & 42.63 & 0.48 & & 77.02 & 0.16 \\ 
   \bottomrule
\end{tabular}
\end{table}

\begin{figure}[ht]
\centering
\begin{minipage}{.3333\textwidth}
	\centering
	\includegraphics[width=.87\linewidth]{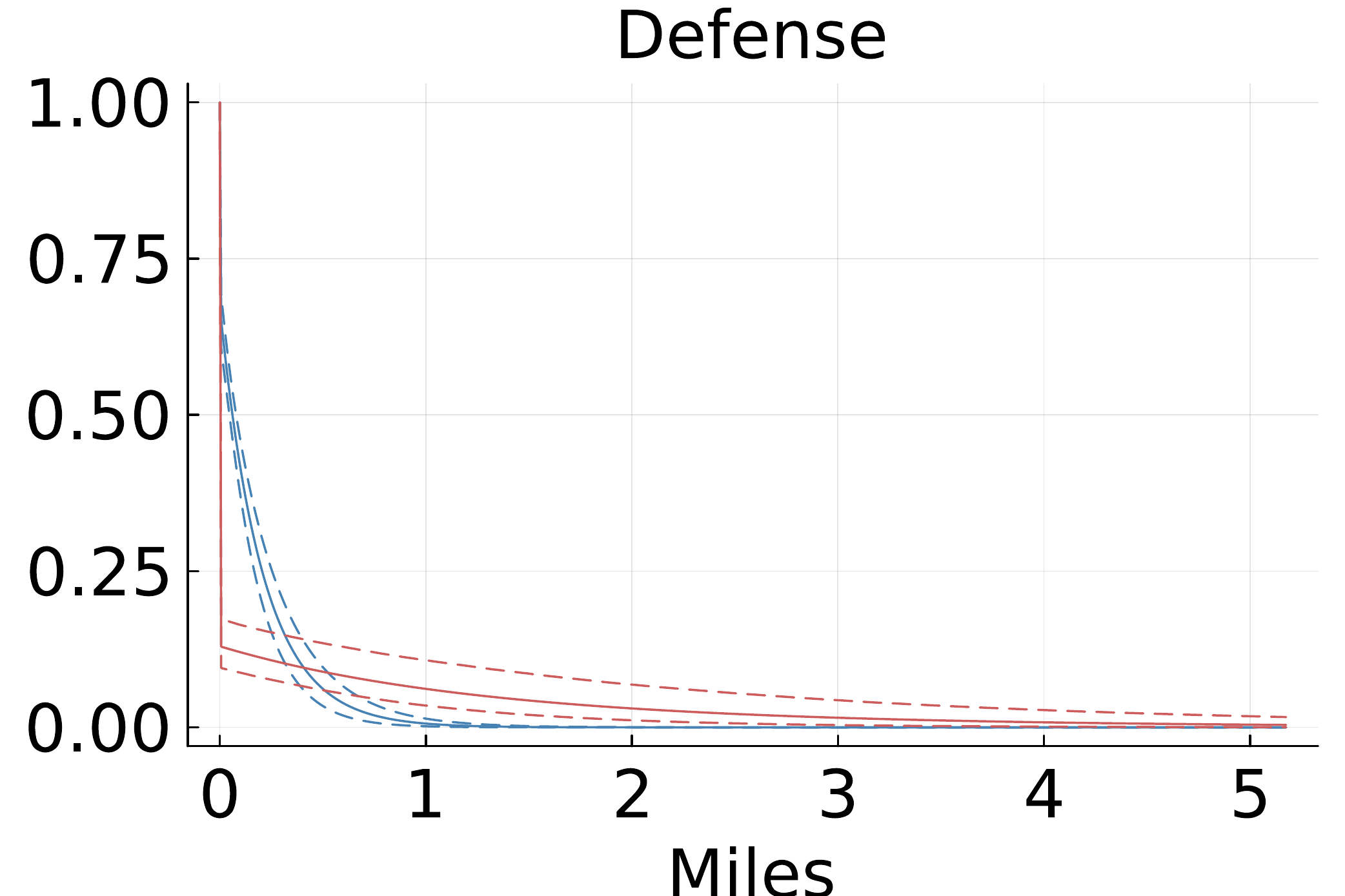}
\end{minipage}%
\begin{minipage}{.3333\textwidth}
	\centering
	\includegraphics[width=.87\linewidth]{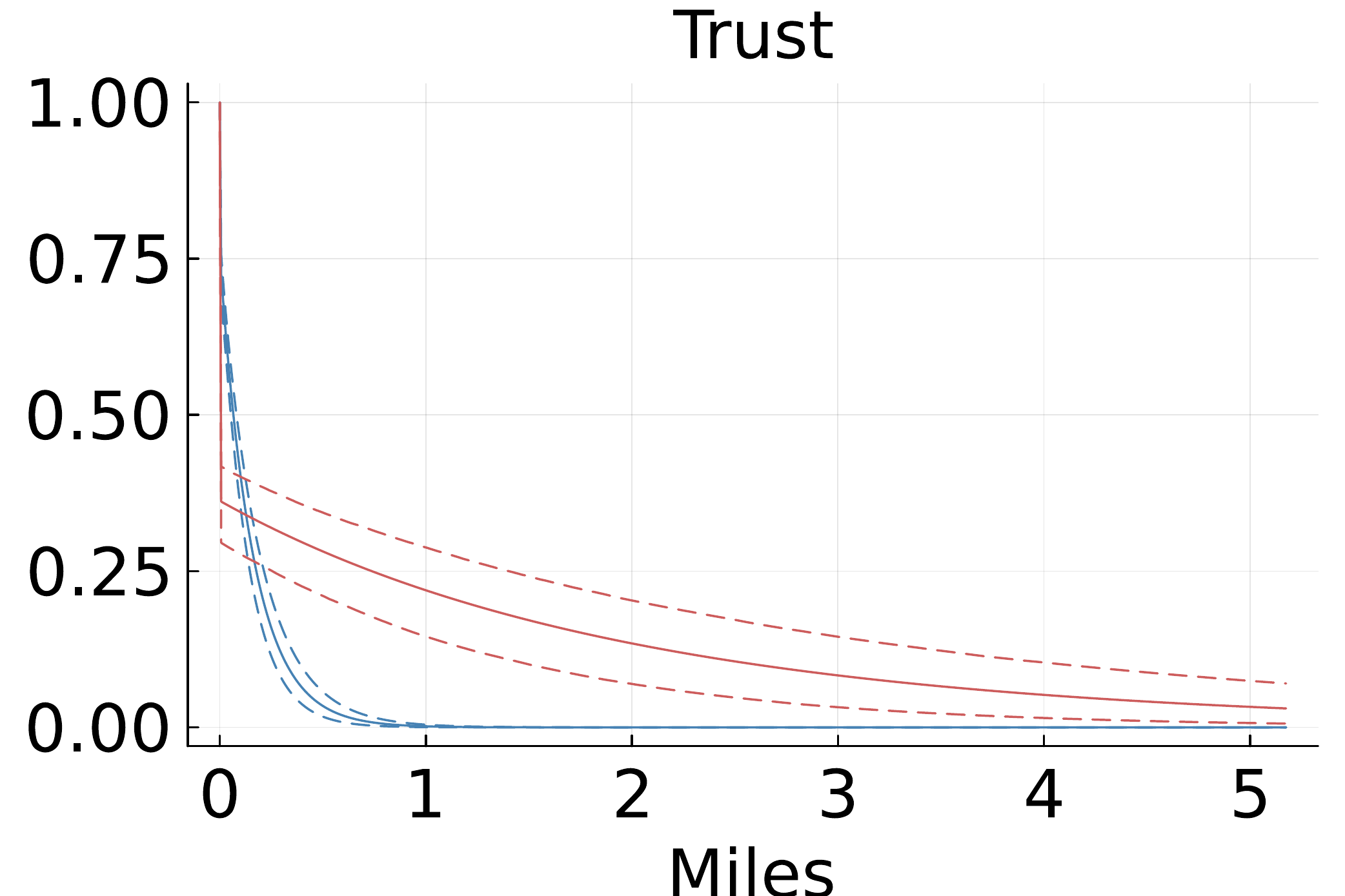}
\end{minipage}%
\begin{minipage}{.3333\textwidth}
	\centering
	\includegraphics[width=.87\linewidth]{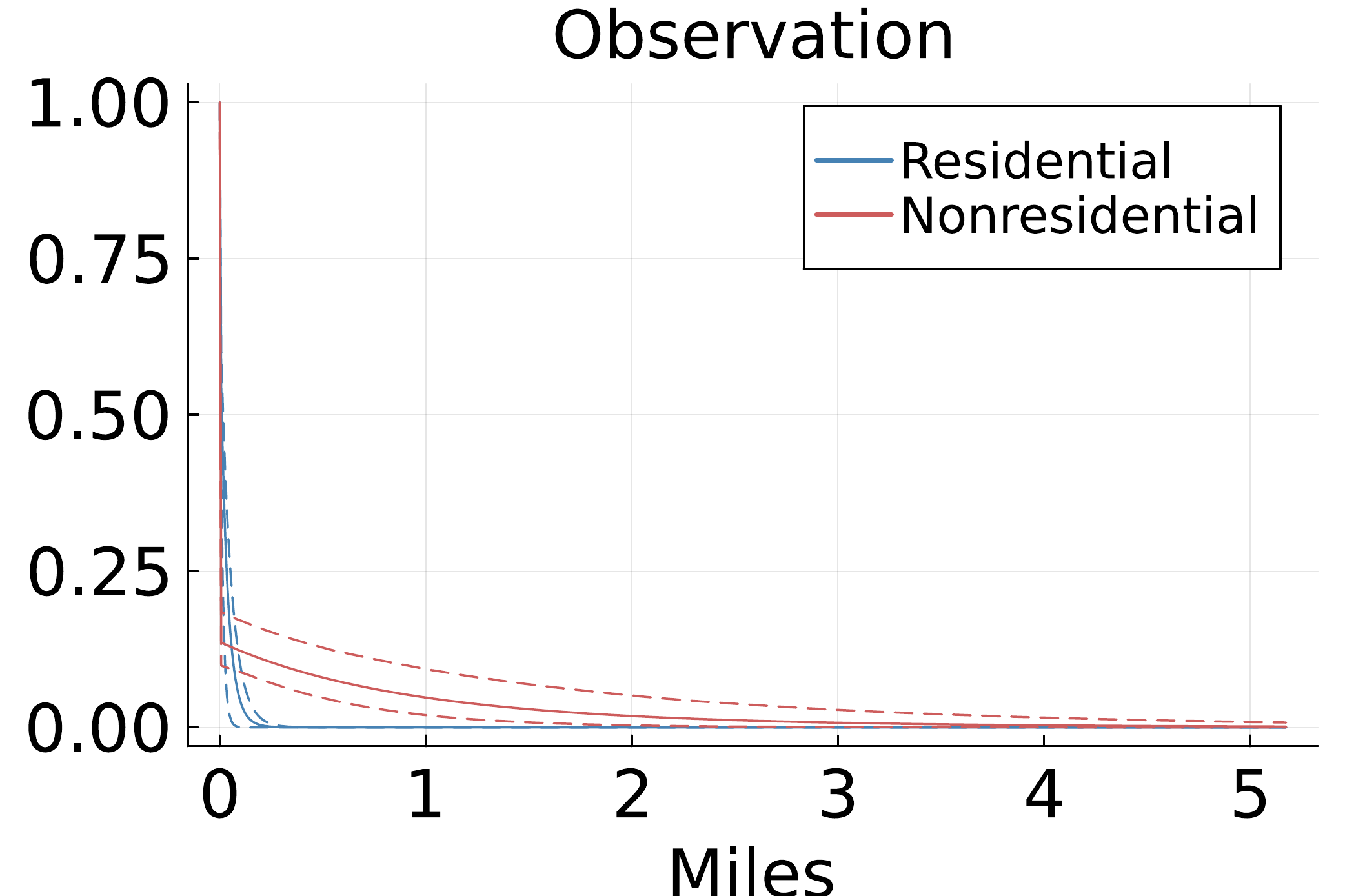}
\end{minipage}
\caption{Estimated posterior mean correlation functions from Bayesian spatial ordinal models
	for residential and nonresidential locations of the defense, trust, and observation component, plotted respectively from left to right. Dashed
	lines indicate lowers and upper bounds of 95\% posterior point-wise credible
	intervals.}
\label{fig:compare_corr_funcs}
\end{figure}

Table~\ref{tab:prelim_model} shows the estimated mean of the marginal posterior distribution of the spatial
range parameter, $\phi$, and spatial proportion-of-variance parameter, $\tau^2$, for all six fitted preliminary models. The
parameter estimates for the models fitted to the residential data differ from the
estimates from the nonresidential data across all three collective efficacy components.
% , with higher estimated values of $\phi$ and $\tau^2$ for residential locations versus nonresidential.
A higher value of $\phi$ indicates
a shorter range of spatial dependence and a high value of $\tau^2$ indicates a
larger proportion of the variance is attributed to the spatial dependence
rather than independent random error. 
% Across all three components, nonresidential locations have longer range spatial
% dependence, but show greater variation within location. 
We plot the estimated posterior mean correlation function with 95\% pointwise credible intervals in Figure~\ref{fig:compare_corr_funcs}. 
For each pair of posterior samples of $\phi$ and $\tau^2$, we calculated the correlation function for a grid of distances from 0 to 5 miles and plot
% \begin{equation}
% \rho(d_{\loc,\loc'};\phi,\tau^2) = 
% \begin{cases}
%     1 &\text{if } d_{\loc,\loc'} = 0 \\
%     \frac{\tau^2}{1+\tau^2}\exp(-\phi * d_{\loc,\loc'}) &\text{if } d_{\loc,\loc'} > 0.
% \end{cases}
% \end{equation}
the pointwise mean and lower/upper bounds of a 95\% credible interval.
For the defense component, the estimated correlation function for the
residential data is characterized by shorter term spatial dependence but a greater
proportion of variation
attributable to spatial dependence. In contrast, the correlation function for
the nonresidential data of the defense component exhibits longer spatial dependence and a much larger proportion of independent error. A similar pattern characterizes the difference between the correlation functions for the trust and observation components.
These differences in the estimated correlation functions between the two types of
locations suggest that a model which accounts for the difference in spatial
dependence between locations of different land-use types will better capture the overall
spatial dependence structure of the data. 

The spatial dependence structure of the model has implications for the main
goal of our analysis: to predict levels of each component of collective
efficacy at unobserved locations. A spatial generalized linear mixed model with a single latent stationary spatial process would most likely lead to a spatial dependence parameter estimate somewhere between the distinct land-use specific spatial dependence parameters
shown in Table~\ref{tab:prelim_model}. We show this is indeed the case in Section~\ref{sec:results}
for all three components of collective efficacy. By expanding the dimension of the response to
include a latent spatial process for each land-use type, we can better separate the distinct spatial dependence structures of each land-use process. This will allow us to have a better
understanding of the spatial process of each component of collective efficacy
at a finer resolution within census tract or block group.   Furthermore, we can
smooth over land-use boundaries through the cross correlations between the latent processes corresponding to land-use
types. 

\section{Land-use filtering model}
\label{sec:methods}
In this section, we introduce a novel land-use filtering model, a Bayesian spatial generalized linear mixed model that incorporates dimension expansion and multivariate spatial modeling strategies. We begin
with a motivating schematic to demonstrate how land-use filtering captures nonstationary behavior in a latent univariate spatial process.  Throughout this section we omit notation references to the three different components of collective efficacy since we will fit the same model to each component separately (see Section \ref{sec:discuss} for more details on the rationale for fitting the components separately). 

\subsection{Motivation}
\label{sec:motivation}
\begin{figure}[htp!]
\centering
\begin{minipage}{.33\textwidth}
  \centering
  \includegraphics[width=.95\linewidth, height=.9\textheight]{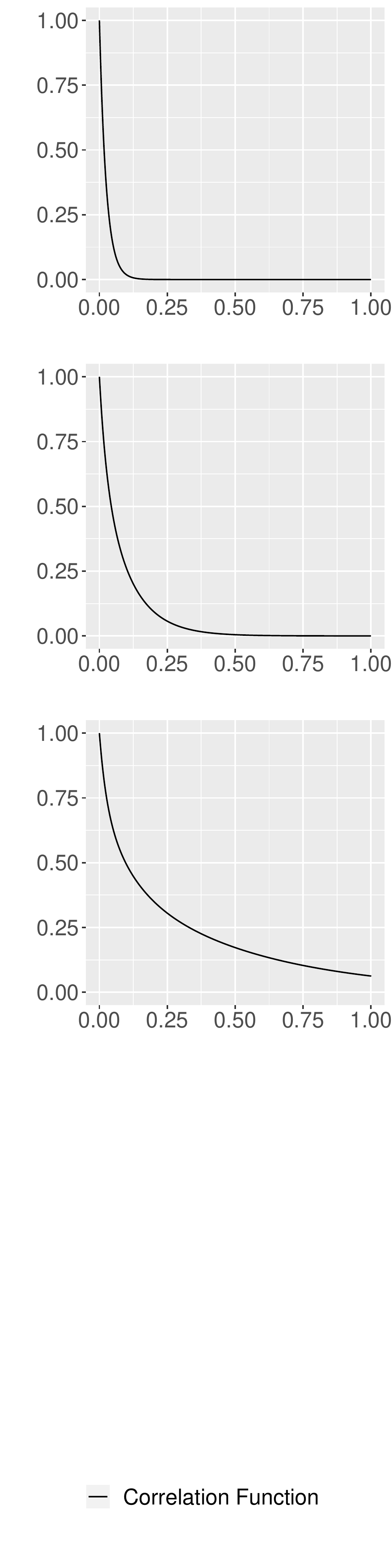}
\end{minipage}%
\begin{minipage}{.33\textwidth}
  \centering
  \includegraphics[width=.95\linewidth, height=.9\textheight]{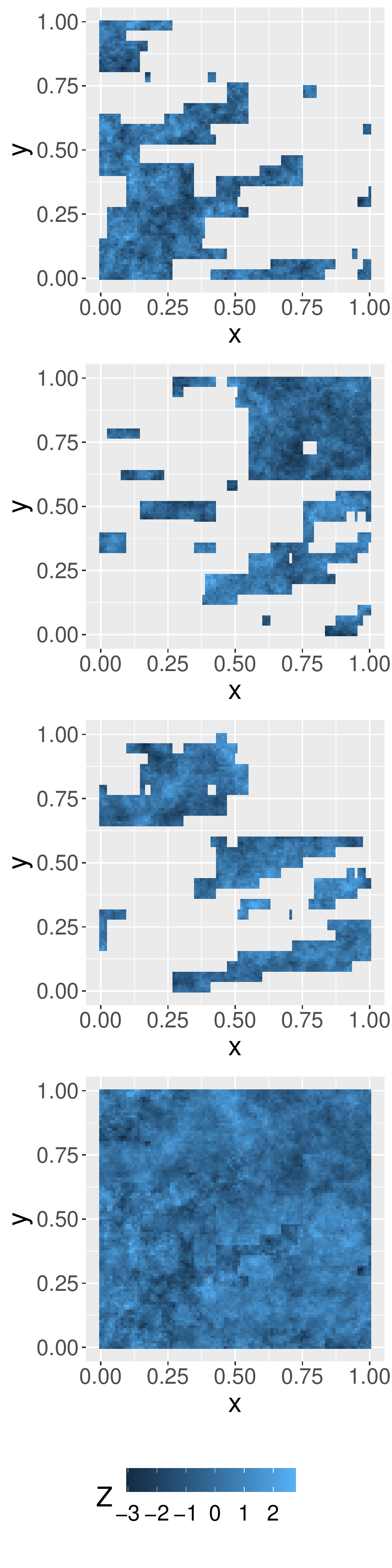}
\end{minipage}%
\begin{minipage}{.33\textwidth}
  \centering
  \includegraphics[width=.95\linewidth, height=.9\textheight]{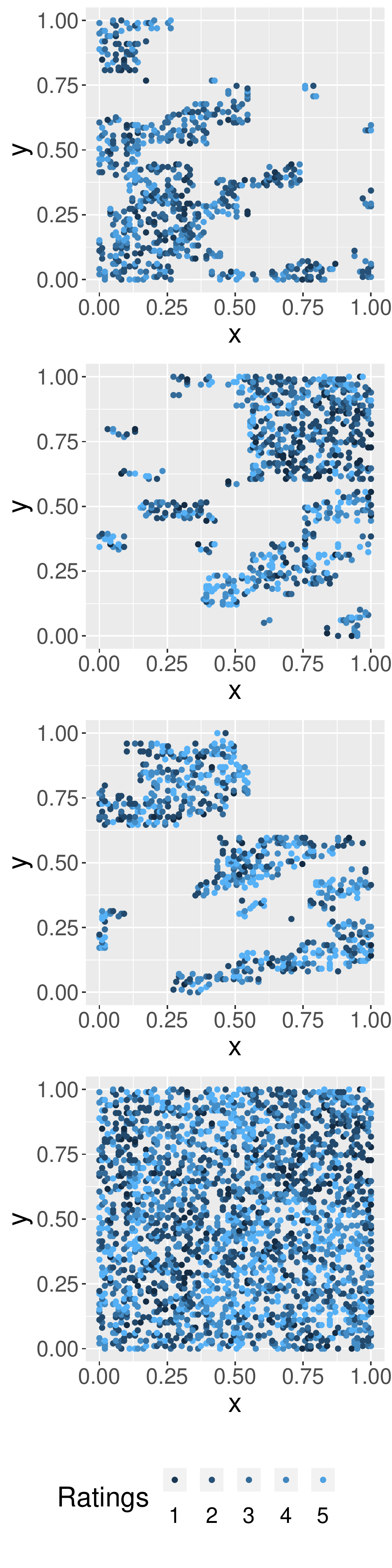}
\end{minipage}
\caption{A demonstration of a generative model for ordinal data under land-use filtering.  See Section~\ref{sec:motivation} for a discussion.}
\label{fig:demo}
\end{figure}

When the study area can be partitioned into different land-use categories (or some other spatial partition), we
can expand the dimension of the latent spatial process in a spatial generalized linear mixed model to allow distinct spatial dependence
structures for each land-use category. To motivate the land-use filtering
model developed formally in the next section, consider the following generative model illustrated in Figure~\ref{fig:demo}.  We generate from a mean-zero latent multivariate spatial process with a known cross-covariance function on a fine grid of locations distributed across a two-dimensional region (a unit square in Figure~\ref{fig:demo}).  The assumed marginal spatial correlation functions for this illustrative example are shown on the left panel.  We then discard the components of each sample that do not correspond to the land-use category of a particular location, producing the top three plots in the center column of Figure~\ref{fig:demo}.  The bottom plot shows the filtered process created by combining the land-use specific component processes at each location.  The resulting image is fairly locally smooth across land-use boundaries due to the cross-component dependence of the latent process, but the spatial dependence structure is not identical across space.  Lastly, we assign an ordinal rating to a set of random locations within the study area based on the value of the latent process at the locations. That is, these ratings are based on the non-discarded component of the simulated random vector associated with that location (top three plots on the right column).  The simulated ratings (bottom right) arising from the filtered latent process (bottom center) are analogous to the AHDC collective efficacy ratings.   
% From the latent process we simulate rankings for
% randomly selected locations by applying cutoffs to obtain ordinal rankings from
% the latent spatial process. The data generating mechanism may be summarized as
% follows.  First draw a multivariate response $\etabf(loc)$ at each location
% $\loc$. We then apply a filter to the multivariate response
% $f(\etabf(\loc),g(\loc)) = \eta_{g(\loc)}(\loc) = Z(\loc)$ where only the
% response for land-use type $q$ of location $\loc$ is recorded. By definition
% the function $g(\loc) \rightarrow q$ maps each location to only one land-use
% type. The ordinal responses are obtained by adding additional independent error
% to the spatial effect and then binning the latent variable into ordered
% categories one through five.

% In Figure~\ref{fig:demo}, the top three panels on the left side show nonoverlapping yet correlated spatial processes for three different land-use
% categories.  In the context of a generative model for ordinal data, these
% represent a latent level of a component of collective efficacy.  The higher the
% latent level, the more likely an individual in that area will rate location
% favorably on the likert scale.  Each spatial process shows different levels of
% smoothness and when all three are plotted together the combined process is
% nonstationary. Each land-use process is assumed to be continuous, however the
% area at which we are able to observe each process does not have to be contiguous.
% An important feature of the model is that each point within the study area can
% be categorized into only a single land-use category. 

Note that in Figure~\ref{fig:demo} when the ratings across
all three land-use categories are plotted together (bottom right column), it is difficult to
visualize different spatial dependence structures which are (more) distinguishable when the latent
processes are plotted together. By expanding the univariate response to allow
for distinct spatial correlation functions for each land-use category and dependence across land-use categories, we can improve model fit for the AHDC collective efficacy ratings. 

\subsection{Land-use filtering ordinal regression}

In specifying our land-use filtering model, we consider only a single component of collective efficacy.  Each of the AHDC caregivers provides their rating on a five-point ordinal scale, where a rating of $K \equiv 5$ is the ``best'' and 1 is the ``worst.''  To facilitate specification of our model, we introduce notation for the ratings, with the primary index indicating the location of the rating.  We let $Y_{ij}$ denote the $j$th rating of the $i$th location, where $\loc_i$ is a unique location in the study area, $\mathcal{S}$, for $i = 1, \dots, m$ and $j = 1, \dots, n_i$.  The constant $m$ is the number of unique locations, and $n_i$ is the total number of caregivers who rated the $i$th location.   We denote the total number of ratings as $n = \sum_{i=1}^m n_i$, and the collection of all ratings as $\mathcal{Y} =
\{Y_{ij}: i=1,\dots,m; j=1,\dots n_i\}$. 

We model the elements of $\mathcal{Y}$ using a Bayesian ordinal probit regression model, specified using the well known data augmentation scheme of \citet{albert_chib1993,
albert_chib1997}, which has been extended to the spatial setting by \citet{oliveira1997, oliveira2000, higgs_hoeting2010, schliep_hoeting2015, berrett_calder2012, berrett_calder2016}. Under this scheme, we introduce continuous latent variables $\mathrm{Z}_{ij}^*$ defined at each of the $m$ locations and for each $n_i$ ratings and unknown cut points, $\cutpt_1, \dots, \cutpt_{K-1}$ such that 
\[
Y_{ij} = 
\begin{cases}
K &\mbox{if } \cutpt_{K-1} < \mathrm{Z}^*_{ij}\\
k &\mbox{if } \cutpt_{k-1} < \mathrm{Z}^*_{ij} < \cutpt_k, \mbox{ for } k = 2, \dots, K-1\\
1 &\mbox{if } \mathrm{Z}^*_{ij} < \cutpt_1
\end{cases}.
\]
Unlike the setting discussed in \citet{higgs_hoeting2010}, we have multiple latent random variables defined at each location due to the fact that some locations are rated by multiple individuals.  Each of these random variables is related to a single latent spatial process $\tilde{Z}(\loc)$ defined for all $\loc \in \mathcal{S}.$  We assume that the $\mathrm{Z}_{ij}^*$s are conditionally independent given the latent spatial process $\tilde Z(\loc_i)$, and parameters $\betabf$, $\thetabf$, and $\sigmabf^2$:
\begin{equation*}
	\mathrm{Z}^*_{ij}|\tilde Z(\loc_i),\betabf,\thetabf,\tilde\sigmabf^2 \sim
	\mathrm{N}(\x_{ij}'\betabf + \tilde Z(\loc_i),\;\tilde\sigma^2_{g(\loc_i)}). \\
\end{equation*}
Here, $\x_{ij}$ is a $p \times 1$ vector of covariates associated with the $j$th
rating of location $i$ and $\betabf$ is the corresponding $p \times 1$ vector of regression coefficients. The vector $\tilde\sigmabf^2$ contains the variance parameters for the independent error of each land-use process and $g(\loc): \mathcal{S} \rightarrow \{1,\dots,Q\}$ is a function that returns the index of the land-use category of location $\loc$.  We use the tilde
notation to highlight the fact that this latent process is constrained so that the total error variance of the $\mathrm{Z}_{ij}^*$s is equal to one. Constraining the latent spatial process in ordinal regression model can aid in the identifiability of the covariance parameters $\thetabf$ \citep{schliep_hoeting2015, yan_etal2007}. We discuss
additional identification considerations for spatial ordinal regression models in
Section~\ref{sec:ident}. 
The vector $\thetabf$ denotes the covariance parameters associated with the spatial process $\tilde Z(\loc)$, on which $\tilde \sigmabf^2$ also depends, due to the constraint on the total error variance of $\mathrm{Z}_{ij}^*$.

Before defining the constrained $\tilde Z(\loc)$ and a model for its unconstrained analogue $Z(\loc)$, we introduce a $Q$-dimensional spatial process $\etabf(\loc)$, where $Q$ is the number of land-use categories in the spatial partition.  We define
$\etabf(\loc)$ to be a multivariate Gaussian process with mean zero and
parametric covariance function $\CovM^{(\etabf)}_{\loc,\loc'}(\thetabf)$ defined by
the linear model of coregionalization (LMC).  The LMC remains the standard approach
to create covariance functions in multivariate spatial modeling
\citep{goulard_voltz1992, grzebyk_wackernagel1994, schmidt_gelfand2003, wackernagel2003,
gelfand_etal2004}. While alternatives to the LMC exist (e.g., \citet{gneiting_etal2010}
and \citet{apanasovich_etal2012} propose Mat\'{e}rn cross-covariance functions for multivariate random fields), \citet{oliveira1997, oliveira2000} has shown that smoothness parameters (such as the
smoothness parameter in the Mat\'{e}rn covariance function) are near
nonidentifiable in spatial ordinal regression models.  Since our $\etabf$ will be a component of a spatial ordinal regression model, we prefer the LMC specification.   

The LMC is a constructive approach for modeling a multivariate spatial process as a linear combination
of independent spatial processes. Let $\w(\loc)$ be a $Q$-dimensional spatial process comprised of independent component processes $\w_1, \dots, \w_Q(\loc)$.  Each $\w_q(\loc)$ is a Gaussian process with mean 0, variance 1, and correlation function $\rho_q(\loc,\loc';\phi_q)$, where $\phi_q$ are unknown parameters.
Again, since smoothness parameters are near nonidentifiable, we specify $\rho_q(\loc,\loc';\phi_q)$ as an exponential correlation function,
\[
\rho_q(\loc,\loc',\phi_q) = \exp(-\phi_q||\s - \s'||),
\]
but note that alternate correlation functions could be used instead.  In the LMC specification, we assume that $\etabf(\loc)=\A\w(\loc)$, where $\A$ is a $Q \times Q$ lower triangular matrix with $Q\times(Q-1)/2$ unknown parameters.  It follows directly that the cross covariance for $\etabf(\loc)$ and
$\etabf(\loc')$ is  
\begin{equation*}
	\CovM^{(\etabf)}_{\loc,\loc'}(\thetabf) = 
		\A\Gammabf(\loc,\loc')\A',
\end{equation*}
where $\thetabf$ consists of the elements of $\phibf$ and unknown elements of $\A$ and $\Gammabf(\loc,\loc')$ is a matrix valued function with
$\rho_q(\loc,\loc',\phi_q)$ on the $q$th diagonal.

We now relate the Q-dimensional process $\etabf(\loc)$ to the univariate process $Z(\loc)$ using land-use filtering (i.e., extracting the component of $\etabf(\loc)$ corresponding to the land-use category of $\loc$).  Formally, we define   
\begin{equation*}
\label{eq:Zmodel}
Z(\loc) = f(\etabf(\loc),g(\loc)) = \eta_{g(\loc)}(\loc) + \delta_{g(\loc)},
\end{equation*}
where $\deltabf = (\delta_1, \dots, \delta_Q)$ is a vector of unknown mean shift parameters defining the mean of the $Z(\loc)$ process.  Writing $Z(\loc) = \abf_{g(\loc)}'\w(\loc)
+ \delta_{g(\loc)}$, where $\abf_{q}$ is the $q$th row of $\A$, it is clear that $Z(\loc)$ is a Gaussian process with mean function $\delta(\loc)$, where
\[
\delta(\loc) = \delta_{g(\loc)},
\]
and cross-covariance function
\begin{equation*} 
\label{eq:LMC_land_filter} 
	\CovM^{(Z)}_{\loc, \loc'}(\thetabf) = 
\abf_{g(\loc)}'\Gammabf(\loc,\loc')\abf_{g(\loc')}.
\end{equation*}
between any two locations $\loc$ and $\loc'$
with land-use types $g(\loc)$ and $g(\loc')$.  Note that $Z(\loc)$ is nonstationary with both a mean function $\deltabf(\loc)$ and  covariance function $\CovM^{(Z)}_{\loc, \loc'}(\thetabf)$ that vary across land-use categories.

%\citep{higgs_hoeting2010}.

% How we specify the distribution of $\mathrm{Z}_{ij}^*$ allows us to account for spatial dependence between ratings at separate locations \citep{higgs_hoeting2010, schliep_hoeting2015,
% berrett_calder2012, berrett_calder2016}. 

To complete the specification of the model, we need to define the covariance function for the constrained $\tilde Z(\loc)$, which allows the $Z^*_{ij}$s to have total variance of one. First we note that the variance of the unconstrained, land-use-category-specific process $\eta_q(\loc) + \delta_q$ is $||\abf_q||^2$.  It follows then that the
constrained covariance is
\begin{equation}
\label{eq:constrain_cov} 
    \tilde\CovM^{(\tilde Z)}_{\loc,\loc'}(\thetabf^*) = \frac{\abf_{g(\loc)}'\Gammabf(\loc,\loc')\abf_{g(\loc')}}{\sqrt{||\abf_{g(\loc)}||^2
	+ \sigma^2_{g(\loc)}}\sqrt{||\abf_{g(\loc')}||^2 + \sigma^2_{g(\loc')}}},
\end{equation}
where $\thetabf^* = \{\mathrm{vec}(\A), \phibf, \sigmabf^2\}$.
Additionally, the independent errors are also constrained so that
\begin{equation*}
\label{eq:constrain_sigma} 
	\tilde\sigma^2_{g(\loc)} = \frac{\sigma^2_{g(\loc)}}{||\abf_{g(\loc)}||^2 +
	\sigma^2_{g(\loc)}}. 
\end{equation*}
The resulting constrained process, $\tilde Z(\loc)$, is a nonstationary Gaussian process
\begin{equation*}
	\tilde Z(\loc)|\delta(\loc),\thetabf^* \sim
	\mathrm{GP}(\delta(\loc),\tilde\CovM^{(\tilde Z)}_{\loc,\loc'}(\thetabf^*)),
\end{equation*}
with mean function, $\deltabf(\loc)$, and covariance function, $\tilde\CovM^{(\tilde Z)}_{\loc, \loc'}(\thetabf^*)$.

% The model outlined above is a spatial generalized linear mixed model (SGLMM)
% \citep{schliep_hoeting2015, berrett_calder2016}. We are required to fit a SGLMM
% as we have replication in our dataset.  For a dataset without replication at
% the location level, one could drop the independent error resulting in a spatial
% generalized linear model \citep{higgs_hoeting2010, schliep_hoeting2015,
% berrett_calder2012, berrett_calder2016}. The model specification above is
% unique as both $Z^*_{ij}$ and $\tilde Z(\loc)$ depend on the covariance
% parameters $\thetabf^*=\{\sigmabf, \thetabf\}$ We use this formulation of the
% SGLMM because restricting the total error variance of $Z_{ij}*$ to one helps
% with the identifiability of $\thetabf^*$ \citet{yan_etal2007}.  

\subsection{Identifiability of model parameters} 
\label{sec:ident}
It is well known that Bayesian probit regression models for ordinal and binary data (and subsequently the spatial extensions of these models) contain nonidentifiable parameters. That is, there is not a one-to-one mapping between the parameters and the value of the likelihood function. We describe below the necessary constraints to obtain likelihood identifiability in the land-use filtering model.

First, a global intercept (or cell means intercepts) and the cut points $\cutpts$ cannot be jointly identified.  To address this identification issue we fix $\delta_1=0$, the mean of the first land-use process, and estimate the remaining $\delta_q$ for $q>1$. Additionally, we exclude a
global intercept from $\betabf$, the vector of regression coefficients.  Alternatively, one could fix one of the cut points, for example
$\cutpt_1 = 0$, which would allow the estimation of a global intercept in $\betabf$
or the estimation of $\delta_1$, but not both.  As a last consideration on the cut points, in a standard multivariate
probit model, one could use a different set of cut points for each
multivariate process, although this would preclude the use of land-use specific intercepts. We opt for a single set of cut points across the land-use
processes so that interpretation of the latent variable $\tilde Z(\loc)$ is
consistent across land-use categories.

In Bayesian spatial probit regression models with an independent error term, the variance of the independent error and the regression coefficients $\betabf$ are identified up to a multiplicative constant. In the hierarchical model specified above, one could drop the constraint on the spatial random effects $\Z(\loc)$ and fix each $\sigma^2_q$ to obtain likelihood identifiability. We instead choose to constrain the total error variance of $\Zrm^*$ to be one, as this can aid the identifiability of the
covariance parameters \citep{schliep_hoeting2015} and facilitates a block update of the covariance parameters in our MCMC algorithm as described in Section~\ref{sec:mcmc}.
Under this scaling regime of equation~\ref{eq:constrain_cov}, we are limited to estimating the proportion of the variance due to spatial dependence, and again the $\sigma_q^2$'s must all be fixed. By fixing $\sigma^2_q=1$ for all $q$, we can express the proportion of variance due to spatial correlation for each land-use category $q$ as $\frac{||\abf_{q}||^2}{||\abf_{q}||^2 + 1}$.

As the caregiver dataset has replication at the location level, we are required to include independent error for the latent process of the ratings. For a setting without replication, one could drop the independent error assumption and let
\begin{equation*}
	\Zrm^*_{i} = \x_{i}'\betabf + \tilde Z(\loc_i).
\end{equation*}
Now we constrain the variance of  $\tilde Z(\loc)$ to be one, as again an unconstrained covariance and $\betabf$ are nonidentified. The constrained covariance for the model without independent error results in the correlation function
\begin{equation*}
	\CorM^{(\tilde Z)}_{\loc,\loc'}(\thetabf)=
	\frac{\abf_{g(\loc)}'\Gammabf(\loc,\loc')\abf_{g(\loc')}}{||\abf_{g(\loc)}||\,
	||\abf_{g(\loc')}||}.
\end{equation*}

Lastly, we discuss another identifiability concern that is uniquely Bayesian, that of partial identifiability. Partially identifiable models are characterized by posterior distributions that do not converge to a point mass as the sample size increases to infinity, yet are still informed by the data and differ from the prior distribution \citep{gustafson2015}. Partially identified models contain partially or weakly identified parameters that are typified by substantive flat regions in the posterior or by a posterior that critically depends on the prior \citep{li_etal2022}. Identifiability, in this context, is on a continuum reflecting the strength of learning from the data. Our model is partially identifiable in that our observed data is a corrupted version of the collective efficacy process that we are trying to predict. In particular, the covariance parameters are weakly identifiable.  As a consequence, we cannot expect to precisely learn about these parameters even when the data are simulated from the model and the sample size is large. Section~\ref{sec:sim_study} investigates the implication of the partial identifiability of our model in terms of out-of-sample predictive performance. 

% %
% \begin{equation}
% %
% 	\tilde Z(\loc)|\thetabf,\delta(\s) &\sim
% 	\mathrm{GP}(\delta(\loc),\CorM^{(\tilde Z)}_{\loc,\loc'}(\thetabf)) \\
% %	
% \end{equation}
% %

% Lastly, at least one of the $\sigma^2_q$ must be fixed for
% identifiability.  When $\A$ is a lower triangular matrix it is intuitive to set
% $\sigma_1^2 = 1$. 
% Note that estimation of the remaining $\sigma^2_q$ for $q>1$ is possible
% because we use a single set of cut points across all $Q$ land-use categories. Allowing a
% different set of cut points for each land-use process would require all of the $\sigma^2_q$s be fixed. For the first observed process, the proportion of spatial
% variation only depends on the first element of $\A$. When $\sigma^2_1=1$, the
% proportion of variance due to the spatial correlation, as opposed to independent
% error, is $\frac{a_{11}^2}{1+a_{11}^2}$, where $a_{qq}$ is the corresponding
% element of $\A$.  When the remaining $\sigma^2_q$, for $q > 1$, are allowed to
% vary this provides greater flexibility in the cross-correlation functions for
% the proportion of independent error of the remaining spatial processes. Estimating
% the $\sigma^2_q$ can be difficult; therefore, we suggest fixing
% $\sigma^2_q=1$ for all $q$. We find that this assumption is not unreasonable as
% the free elements of the matrix $\A$ provide enough flexibility to account for
% varying amounts of independent error for each land-use process.

\subsection{Implications of the LMC specification}
\label{sec:lmc_implications}
We now explore the implications of the LMC specification on the cross-covariance function of the multivariate latent process.
% The LMC is prevalent because it is a straightforward and intuitive way to
% obtain positive definite covariance matrices for a multivariate response: each
% component of the multivariate process is defined as a linear combination of independent
% univariate spatial processes. While the method is intuitive, it does impose
% constraints on the flexibility of the covariance and cross-covariance
% functions. 
In the continuous response setting, \citet{gneiting_etal2010} note that when all elements of the matrix $\A$ are treated as unknown parameters (i.e. no
structural zeros), the smoothness of each latent process is determined by the roughest latent component and that by fixing the upper right elements of the matrix $\A$ to be zero,
distinct smoothness properties can be estimated. In a similar vein, we find that the structural zeros also help in identifying distinct spatial range parameters for each component process. 

Imposing this structure of $\A$, however, does indirectly place restrictions on the cross-covariance function.  Consider the setting where the dimension of the latent process, $Q$, is three and $\A$ is  lower triangular such that $a_{qq'}=0$ for all $q'>q$, where $a_{qq'}$ denotes the $(q,q')$ element of $\A$.  Let $\tilde \etabf(\loc)$ be the constrained multivariate process defined by the LMC with the constraint given in Equation~\ref{eq:constrain_cov}.
%The cross-covariance matrix-valued function for $\tilde \etabf(\loc)$ as each element of the matrix can appear in the cross-covariance function for the  $\tilde Z(\loc)$ across land-use types and locations (again see Equation~\ref{eq:constrain_cov}).  
The cross-covariance matrix-valued function 
for $\tilde \etabf(\loc)$ and $\tilde \etabf(\loc')$ can be written as
\begin{equation}
\label{eq:cross_corr_matrix}
\begin{bmatrix}
\frac{a_{11}^2 \rho_1}{||\abf_1||^2 + \sigma^2_1} &
\frac{a_{11}a_{21} \rho_1}{\sqrt{||\abf_1||^2 + \sigma^2_1}\sqrt{||\abf_2||^2 +
	\sigma^2_2}} &
\frac{a_{11}a_{31} \rho_1}{\sqrt{||\abf_1||^2 + \sigma^2_1}\sqrt{||\abf_3||^2 +
	\sigma^2_3}}  \\
\frac{a_{11}a_{21} \rho_1}{\sqrt{||\abf_1||^2 + \sigma^2_1}\sqrt{||\abf_2||^2 +
	\sigma^2_2}} &
\frac{a_{21}^2\rho_1 + a_{22}^2\rho_2}{||\abf_2||^2 + \sigma^2_2} &
\frac{a_{21}a_{31}\rho_1 + a_{22}a_{32}\rho_2}{\sqrt{||\abf_1||^2 +
	\sigma^2_1}\sqrt{||\abf_3||^2 + \sigma^2_3}} \\
\frac{a_{11}a_{31} \rho_1}{\sqrt{||\abf_1||^2 + \sigma^2_1}\sqrt{||\abf_3||^2 +
	\sigma^2_3}} &
\frac{a_{21}a_{31}\rho_1 + a_{22}a_{32}\rho_2}{\sqrt{||\abf_1||^2 +
	\sigma^2_1}\sqrt{||\abf_3||^2 + \sigma^2_3}} &
\frac{a_{31}^2\rho_1 + a_{32}^2\rho_2 + a_{33}^2\rho_3}{||\abf_3||^2 +
	\sigma^2_3}
\end{bmatrix},
\end{equation}
where $\rho_q=\rho(\loc, \loc', \phi_q)$.  
From this expression for the cross-correlation function in Equation~\ref{eq:cross_corr_matrix}, it is clear that the cross-covariance between the first latent process and the other two only
depends on $\rho_1$. That is, the structural zeros in $\A$ restrict the cross-covariance function between the first and
subsequent latent process to only depend on the spatial dependence structure of the first spatial process. It immediately follows that the lower-triangular form of $\A$ implies that the ordering of the land-use categories will affect the estimated cross-covariance function. 
% Selecting the
% land-use category with shorter spatial range as first in the LMC specification spatial
% dependence between the first land-use category and subsequent land-use categories. In contrast,
% having a land-use process with longer range dependence implies long range spatial
% dependence for cross correlations between the first and subsequent categories.
Additionally, we note that ordering of the components
can affect the (partial) identifiability of distinct spatial range parameters of
the spatial processes.  We discuss our strategy for ordering land-use categories in Section~\ref{sec:results}. 

Additionally, the implied cross-covariance function from the LMC model also
has implications for the trade off between independent error and spatial variance when all
$\sigma_q^2$s are set to one. For each land-use process the proportion of
independent error is $\frac{1}{||\abf_q||^2+1}$.  A land-use process with greater
independent error will have proportionally less spatial error. Because the total error
variance is constrained to one, the land-use processes $\eta_{q}(\loc) + \delta_q$ will have smaller
spatial variance for land-use types with a greater proportion of
independent error. 

\subsection{Model fitting}
\label{sec:model_fitting}
The full Bayesian model is specified by placing priors on the covariance
parameters $\thetabf^*$, and mean function
parameters $\deltabf$ and $\betabf$. A normal prior for $\deltabf$ and
$\betabf$ is conditionally conjugate in the data augmented probit model. 
The diagonal elements of $\A$,
$\phibf$, and $\sigmabf^2$ have support on the positive real line, while the off diagonal elements of $\A$ have support
on the real line. 

We wrote a custom MCMC algorthim in \texttt{Julia} to fit the model \citep{julia2017, jl_differentialequations2017, jl_distributions2019, jl_distributions2021, jl_fftw2005, jl_optim2018}.  As MCMC is
computationally time consuming and burdensome we also propose an approximation
method that does not utilize the data augmentation framework, but rather treats
the ordinal ratings as a Gaussian outcome. We fit the approximate model by direct
maximization of the approximate posterior. We say ``approximate'' as it is not appropriate to assume ordinal data follow a normal distribution, but for the AHDC collective efficacy data spatial predictions from this approximate model are nearly identical to those produced by the land-use filtering model and can be obtained much faster (computation time is hours instead of days).

To facilitate the discussion of each method used to fit the model we introduce
vector notation. Let $\Y$ be a $n\times 1$ vector of ratings and $\Z^*$ be the
$n \times 1$ latent response vector with $Y_r$ and $\Zrm^*_r$ denoting the $r$th rating and latent variable. The vector $\tilde \Z=( \tilde
\Zrm_1,\dots,\tilde \Zrm_m)'$ contains the spatial random effects for each
location, where $\tilde \Zrm_i=\tilde Z(\loc_i)$. The $r$th row of the design matrix, $\X$, contains the covariates $\x_r$ of the $r$th rating.  $\HH$ is a $n\times m$
matrix of ones and zeros that associates the spatial random effects to the
corresponding locations of each rating. The $n \times 1$ vector $\epsilonbf$ of
independent errors have a unique constrained variance, $\tilde\sigma_q^2$, for
each land-use type $q$. We then write the model as
\begin{equation*}
	\Z^* = \X\betabf+\HH\tilde\Z+\tilde\epsilonbf,
\end{equation*}
where vector of spatial random effects, $\tilde \Z$, has a multivariate normal distribution,
\begin{equation*}
\tilde\Z|\deltabf,\thetabf^* \sim \mathrm{MVN}(\M\deltabf, \tilde\CovM(\thetabf^*)),
\end{equation*}
and $\M$ is a $m \times Q$ matrix of zeros and ones that associates the $i$th
location with the mean shift $\delta_q$ of the corresponding land-use type. Lastly, the $\tilde \epsilonbf$ are independent,
\begin{equation*}
\tilde\epsilonbf|\thetabf^* \sim \mathrm{N}(\0, \tilde\D),
\end{equation*}
with $\tilde\sigma_q^2$ on the $r$th diagonal of $\tilde\D$ for
rating $r$ with land-use type $q$.
 
\subsubsection{MCMC}
\label{sec:mcmc}
To facilitate mixing of the MCMC algorithm, we integrate out the spatial random effects.  After doing so, we write the model as
\begin{equation*}
	\Z^* = \X\betabf + \M^*\deltabf  + \nubf\\,
\end{equation*}
with the error term $\nubf$ having a multivariate normal distribution,
\begin{equation*}
	\nubf | \thetabf^* \sim \mathrm{N}(0, \CorM(\thetabf^*)),
\end{equation*}
where $\M^* = \HH\M$ and
\begin{equation}
	\label{eq:vec_sglmm}
\CorM(\thetabf^*)=\HH \tilde\CovM(\thetabf^*)\HH' + \tilde\D
\end{equation}
is a correlation matrix.  First, we update the cut points and latent vector $\Z^*$.  The cut
points are drawn from a uniform distribution $\gamma_k \sim
\mathrm{U}(\max(\Zrm_r^*|Y_r = k),\min(\Zrm_r^*|Y_r > k))$. While alternative schemes have been proposed for updating the cut points, these strategies prove to be impractical computationally for our relatively large data set in the spatial setting \citep{albert_chib1997, cowles1996, higgs_hoeting2010}. Each $\Zrm_r^*$ in the
latent vector $\Z^*$ is drawn from a normal distribution obtained by
conditioning on $\betabf$, $\deltabf$ and all other observations $\Z^*_{-r}$ and
truncated to support $[\gamma_{Y_{r-1}},\gamma_{Y_r}]$. As the precision
matrix, $\Lambdabf = \CorM(\thetabf^*)^{-1}$, is used to evaluate the likelihood
of $\Z^*$ in the posterior for a Metropolis-Hastings update of $\thetabf$, we
also write the variance of
$\Zrm_r^*|\Z_{-r}^*$ in terms of the precision to avoid additional inverse calculations:
$\mathrm{var}(\Zrm_r^*|\Z^*_{-r})=1/\Lambdabf_{rr}$.  After obtaining a draw of $\Z^*$,
the regression coefficients $\betabf$ and mean shift $\deltabf$ can be drawn
jointly from their full conditional distributions with Gibbs updates.  We update $\thetabf$
jointly with a Metropolis-Hastings step evaluating the posterior conditional on
the current draw of $\Z^*$. We use a multivariate normal proposal distribution
for $\thetabf$ centered around the current draw of $\thetabf$. We tune the
covariance matrix for the proposal by running the algorithm to obtain at least
ten thousand draws with a diagonal covariance. From this initial run, we scale
the covariance of the draws of $\thetabf$ by $0.25$ and use the resulting matrix as the
covariance of our proposal distribution for the final run of the MCMC. 

\subsubsection{Approximate maximum a posteriori estimation} 
\label{sec:amap}
While running the
full MCMC algorithm allows us to assess uncertainty in the model parameters, the algorithm is very
time consuming to run due to the matrix inversions required to compute the likelihood
of the latent vector $\Z^*$.  As an alternative, we consider a strategy to estimate the model
parameters that finds the maximum \textit{a posteriori} estimates of parameters from a model that approximates the full Bayesian ordinal specification.
 The approximate model treats the outcomes as Gaussian, so that 
\begin{align*}
	\Y^* &= \X\betabf + \M^*\deltabf  + \nubf\\
	\nubf | \thetabf^* &\sim \mathrm{N}(\0, \CovM(\thetabf^*)),
\end{align*}
where $\CovM(\thetabf^*) = \HH\CovM(\thetabf)\HH' + \D$.  The main difference 
between the ordinal model and the approximate model
is that the covariance of $\nubf$ is unconstrained. To simplify the calculations required to find the \textit{a posteriori} estimates under this approximate model,
we integrate out $\betabf$ and $\deltabf$ and only estimate the covariance
parameters $\thetabf^*$. The marginal distribution of $\Y$ is then given by
$$ \Y|\thetabf^* \sim \mathrm{N}(\0,\X\CovM_{\betabf}\X' +
\M^*\CovM_{\deltabf}{\M^*}'+\HH\CovM(\thetabf)\HH' + \D), $$
where $\CovM_{\deltabf}$ and $\CovM_{\betabf}$ are the covariance from the priors
placed on $\deltabf$ and $\betabf$. We optimize for $\thetabf^*$ by placing
priors on $\thetabf^*$ and then finding the maximum \textit{a posteriori} estimate of
the approximate marginal posterior. 

\subsection{Prediction of the latent collective efficacy process}
While the land-use filtering model is built on the
assumption that we only observe a single land-use process at each location,
when making predictions we can predict the latent level of our response for
all land-use categories at any location. These predictions may be useful in
visualizing how the spatial processes differ from one another, however,
predictions of land-use processes at locations that do not match the true
land-use category may not be meaningful depending on the application at hand.
It would not be correct, for our data example, to predict the change in the
latent level of defense if a location changed land-use categories from residential
to nonresidential.

We now detail how to obtain
samples from the posterior predictive distribution for the spatial process
$\tilde Z(\loc)$ at any location.
Let $\bar\Scal$ be the ordered set of new locations at which we want to
obtain predictions of the spatial process $\tilde Z(\loc)$  for $\loc \in \bar\Scal$ for a component of
collective efficacy. Let $\bar\M$ be the matrix which associates prediction
locations to a land-use category. Additionally, let draws from the posterior distribution
be indexed by $b$. To obtain a draw $\tilde \Z^{[b]}$ from the posterior
predictive distribution we need to evaluate the following covariance matrices plugging in the posterior samples of model parameters (denoted by superscript $[b])$:
$\tilde\CovM_{\bar\Scal,\Scal}(\thetabf^{*[b]})$ and
$\tilde\CovM_{\bar\Scal,\bar\Scal}(\thetabf^{*[b]})$ from
Equation~\ref{eq:constrain_cov} and $\CorM_{\Scal,\Scal}(\thetabf^{*[b]})$ from
Equation~\ref{eq:vec_sglmm}. Then using properties of the multivariate normal
distribution, we draw $\tilde \Z^{[b]}$ from $\tilde \Z|\Z^{*[b]}$ from $\mathrm{N}(\mubf^{[b]},
\CovM^{[b]})$, where
\begin{equation}
\begin{aligned}
	\label{eq:post_spat_re} 
\mubf^{[b]} &=  \bar\M\deltabf^{[b]} + \tilde\CovM_{\bar\Scal,\Scal}(\thetabf^{*[b]})
	\CorM_{\Scal,\Scal}^{-1}(\thetabf^{*[b]}) (\Z^{*[b]}- \X\betabf^{[b]} - \M^*\deltabf^{[b]})
	\\
\CovM^{[b]} &= \tilde\CovM_{\bar\Scal,\bar\Scal} (\thetabf^{*[b]}) -
	\tilde\CovM_{\bar\Scal,\Scal}(\thetabf^{*[b]})
	\CorM_{\Scal,\Scal}^{-1}(\thetabf^{*[b]})
	\tilde\CovM_{\Scal,\bar\Scal}(\thetabf^{*[b]}).
\end{aligned}
\end{equation}
The $\tilde \Z^{[b]}$s are samples from the posterior predictive distribution $\tilde \Z \vert \mathcal{Y}$.

\section{Analysis of the AHDC caregiver data}
\label{sec:results} 
We fit our proposed land-use filtering ordinal regression model to the caregiver ratings, separately for each of the three
components of collective efficacy. As noted in Section~\ref{sec:lmc_implications}, the ordering of the land-use processes in the LMC specification is consequential for identifiability of distinct spatial range parameters and has implications on the cross-covariance between land-use types. From our preliminary analyses in Section~\ref{sec:prelim_model}, we found that the residential processes had shorter spatial range than the nonresidential latent processes. Therefore, we selected residential as the first component in the LMC specification of the multivariate process $\etabf(\loc)$ and nonresidential as the second. We then set the ``other'' category third,
as the ``other'' category has very few observations ($\approx 50$) for
all three components of collective efficacy.   Moreover, we set the third latent component process in the LMC
specification for the covariance function
to be spatially
independent (i.e. setting $\phi_3$ to positive infinity) as we found that 
the range parameter of the third latent process, $\phi_3$,
 is weakly identifiable for each component and MCMC chains did not appear to converge when we attempted to estimate it.  
% Setting $\phi_3$ to positive infinity has the effect of 
% adding an additional independent error to the latent processes of the ``other'' category.
As we expect the components of collective efficacy to differ in terms of the nature of the spatial dependence structure between
nonresidential and ``other'' land-use areas and the ``other'' category
covers a nonnegligible portion of the city, we could not justify collapsing the nonresidential and ``other'' categories into a single land-use category.   

In the design matrix, $\X$, we include
time-of-day indicators (daytime, nighttime, or mixed) and the
day-of-week indicators (weekday only, weekend only, or mixed) as reported by the
survey participants.  We allow the fixed effects corresponding to time-of-day and day-of-the-week to differ by land-use category.
The baseline category is residential locations rated during the daytime. We do not include a day-of-week effect for residential locations as nearly all reports were day-of-week mixed. For the ``other'' category, we collapse day-of--week categories weekend-only and mixed and time-of-day categories night-only and mixed due to data availability.  

We placed $\norm(0,1)$
priors on each of the $\beta$'s and $\delta$'s,  independent $\norm(0,100)$ priors on the off diagonal elements of $\A$, $\mathrm{Cauchy}^+(0,1)$ on each of the range parameters
$\phi_i$ and diagonal elements of $\A$, and a flat prior on the cut points. 
Ordering of the cut points in the posterior is imposed by the likelihood of the latent variables.
For each land-use filtering model, we ran each MCMC for 200,000 interations discarding the first 100,000 as burn in. 

\subsection{Model comparison}
We compare the land-use filtering model for each collective efficacy component to a model that assumes a single stationary latent process. The only difference in the models is the specification of
the correlation matrix. As such, we use the same $\norm(0,1)$
priors on each of the $\beta$'s and $\delta$'s and a flat prior on the cut points. While the land-use filtering model defines the correlation matrix, $\CorM(\thetabf^*)$, as in Equation~\ref{eq:vec_sglmm},  the stationary model defines the correlation as
$\CorM(\bar\thetabf) = (1-\kappa)\CorM(\phi) + \kappa \I$, where $\kappa= 1/(1+\tau^2)$, $\I$ is the identity matrix, $\CorM(\phi)$ is
defined using the exponential correlation function and $\bar\thetabf=\{\phi, \tau^2\}$. We place $\mathrm{Cauchy}^+(0,1)$ priors on $\phi$ and $\tau^2$. As we are not interested in using predictions from the stationary model and the covariance parameters (i.e. only $\phi$ and $\tau^2$) converge more quickly, we obtain 10,000 samples and discard the first 2,000 as burn in.

To compare the fitted models, we use the Widely Applicable
Information Criterion (WAIC) \citep{watanabe2010}.
Information criteria use the log predictive density and include a penalty for over
fitting. In our models, the predictive density (or likelihood) for the data augmentation model
is recovered by integrating out the latent variable $\Z^*$, resulting in an
integral over the multivariate normal distribution:
\begin{equation*}
\label{eq:mult_norm_int}
p(\Y|\thetabf^*, \deltabf, \betabf, \gammabf) = \int \dots \int
	p(\Y,\Z^*|\thetabf^*, \deltabf, \betabf, \gammabf)d\Z^*. 
\end{equation*}
%
%
% Unlike the Deviance Information Criterion (DIC)
% \citep{spiegelhalter_etal2002}, the 
WAIC is defined on the log pointwise posterior predictive density, which  requires the posterior predictive
density to be factored by individual observations. In the land-use filtered model this
factorization is obtained by conditioning on the spatial random effects
$\tilde\Z$ in the model,
\begin{equation*}
\begin{aligned}
p(\Y|\thetabf^*, \tilde\Z, \betabf, \gammabf) &= \int \dots \int
	p(\Y,\Z^*|\thetabf^*, \tilde\Z, \betabf, \gammabf)d\Z^* \\
&= \int_{C_1} p(Z_{1}^*|\thetabf^*, \tilde\Z, \betabf, \gammabf)d Z_1^* \dots
	\int_{C_n} p(Z_{n}^*|\thetabf^*, \tilde\Z, \betabf, \gammabf) dZ_n^* \\
	&= \prod_{r=1}^n [\Phi(\gamma_{Y_{r}} - \x_r'\betabf - \hh_r'\tilde \Z) ) -
	\Phi(\gamma_{Y_{r} -1 } - \x_r'\betabf - \hh_r'\tilde \Z )],
\end{aligned}
\end{equation*}
where $\Phi()$ is the standard normal cumulative distribution function and $\hh_r$ is the column vector of the $r$th row of $\HH$. The distribution of $Y_r$ given $\Zrm^*_r$ is simply the indicator function $\mathrm{I}(\cutpts_{Y_r-1}<\Zrm^*_r<\cutpts_{Y_r})$, which gives the above bounds of integration 
$$
C_r = [\gamma_{Y_r - 1}, \gamma_{Y_r}],
$$
where we define $\cutpt_0=-\infty$ and $\cutpt_K=\infty$.
% \todo{Brandon:  Check equation 14.  Why don't the $Y$s and the $Z^*$s have two subscripts?}. 
As a post processing step of the MCMC output, we obtain samples from the
posterior distribution of the spatial random effects by drawing samples from
the full conditional distribution given in Equation~\ref{eq:post_spat_re} for
all observed locations.

% Conditioning on the spatial random effects not only
% allows us to calculate WAIC, but also allows calculation of DIC without having
% to compute the multivariate normal integral in Equation~\ref{eq:mult_norm_int}.
% WAIC for a model without independent error cannot be computed and DIC must be computed
% using the predictive distribution defined in Equation~\ref{eq:mult_norm_int}. \todo{Brandon:  I think this is confusing to talk about the DIC.  Do you need to?}

\subsection{Results}
We begin with a comparison of model fit using WAIC. The WAIC of the land-use filtering model and stationary model for each component of collective efficacy is given in Table~\ref{tab:waic}. To calculate WAIC we drew the spatial random effect from thinned (every 8th draw after burn in for the stationary models and every 100th draw after burn in for the land-use filtering models) MCMC chains to give us 1000 draws from each model. Thinning facilitated a quicker computation of WAIC.
For all three components, the WAIC is lower for the land-use filtering model, indicating a better model fit. The dimension expansion technique applied in the land-use filtering model allows us to estimate different spatial processes for each land-use type, allowing for a better fit of the underlying spatial process for each component of collective efficacy across land-use types. 

\begin{table}[ht]
\caption{WAIC scores for the land-use filtering and the stationary model}
\label{tab:waic}
\centering
\begin{tabular}{rrrrrr}
  \toprule
  & \multicolumn{1}{c}{Defense} & & \multicolumn{1}{c}{Trust} & & \multicolumn{1}{c}{Observation} \\
 \cmidrule{2-2} \cmidrule{4-4} \cmidrule{6-6}
 Land-use Filtering & 17013.3 && 13203.9 && 17998.1 \\
 Stationary Covariance & 17732.3 && 13968.6 && 18871.0 \\
 \bottomrule
\end{tabular}
\end{table}

Now we discuss how parameter estimates from our land-use filtering and stationary models compare, noting that these comparisons should be viewed heuristically given the partial identifiability of the covariance parameters characterizing the strength of spatial dependence.
Tables~\ref{tab:land-use_filter} and \ref{tab:stationary} give the posterior mean estimates of the parameters in the land-use filtering and stationary models respectively.
% The land-use filtering processes for each of the components of collective efficacy, trust, defence, and observation have a similar characterization. The residential processes (which depends only on $\phi_1$ and $a_{11}$) are characterized by short range spatial dependence and a small proportion of independent error. For the observation process, $\phi_1$ is sufficiently high that the residential ratings are practically independent. The nonresidential processes (which depends on the combination of $\phi_1$, $\phi_2$ through $a_{21}$, and $a_{22}$ have much longer range spatial dependence due to the greater weight $a_{22}$ on the longer range $\phi_2$. Again, the observation component of collective efficacy estimates less spatial dependence for nonresidential locations as compared to nonresidential locations for trust and defense. For observation, $a_{21}$ is estimated to be near zero, which indicates that the nonresidential land-use process is independent from the residential process. By separating the latent processes of the ratings into land-use categories we are better able to estimate how the processes vary within and across land-use categories. For the ``other'' land-use processes, the estimates of the weights are highly variable due to the very few observations in the third category. The weight $a_{33}$ of the third component is near zero for all three components of collective efficacy, indicating that additional independent error is unnecessary for model fit. 
For the stationary models, the single estimate of $\phi$ is in between estimates of $\phi_1$ and $\phi_2$ from the land-use filtering model. The stationary model smooths evenly over the entire study region and masks variation by land-use category.

\begin{table}[ht]
 \setlength\tabcolsep{4pt}
\caption{Posterior mean estimates and corresponding posterior 95\% credible intervals of the parameters in the land-use filtered Bayesian ordinal regression model. The numeric subscripts for the $\beta$'s and $\delta$'s correspond to the land-use processes: 1 for residential, 2 for nonresidential, 3 for the ``other'' category. Note, with the LMC, the subscripts for the $a_{qq'}$'s correspond to the linear weights that relate the independent component processes $w_q(\loc)$, each with spatial range $\phi_q$, to the multivariate outcome $\etabf(\loc)$. For example, the residential process is defined by $a_{11}$ and $\phi_1$, the nonresidential process is defined by $a_{21}$, $a_{22}$, $\phi_1$ and $\phi_2$. The third component process $w_3(\loc)$ is spatially independent, thus $\phi_3$ is fixed at positive infinity. The subscripts ``tod'' and ``dow'' stand for time of day and day of week with ``n'' for \textbf{n}ight, ``m'' for \textbf{m}ixed, and ``e'' for week\textbf{e}nd.}
\label{tab:land-use_filter}
	\centering
\begin{tabular}{lrr@{\hskip 1pt}rrrr@{\hskip 1pt}rrrr@{\hskip 1pt}r}
	\toprule[1.5pt]
	& \multicolumn{3}{c}{Defense} &	& \multicolumn{3}{c}{Trust} & & \multicolumn{3}{c}{Observation} \\
	\cmidrule{2-4} \cmidrule{6-8} \cmidrule{10-12}
	& \shortstack[r]{Posterior \\ mean} & \multicolumn{2}{c}{\shortstack{95\% Credible \\ interval}} 
	& & \shortstack{Posterior \\ mean} & \multicolumn{2}{c}{\shortstack{95\% Credible \\ interval}} 
	& & \shortstack{Posterior \\ mean} & \multicolumn{2}{c}{\shortstack{95\% Credible \\ interval}} \\
  \midrule
	$\phi_1$ & 325.06 & (259, & 408) & & 405.40 & (333, & 486) & &  2142.49 & (1215, & 4797) \\
    $\phi_2$ & 2.54 & (0.43, & 7.88) & & 24.29 & (11.5, & 41.7) &  & 92.70 & (58.6, & 136.0) \\ 
	\midrule
    $a_{11}$ & 1.39 & (1.22, & 1.56) &  & 1.76 & (1.61, & 1.92) &  & 1.60 & (1.48, & 1.73) \\ 
    $a_{21}$ & 0.32 & (0.22, & 0.41) &  & 0.31 & (0.18, & 0.43) &  & 0.05 & (-0.12, & 0.20) \\ 
    $a_{22}$ & 0.58 & (0.42, & 0.72) &  & 0.68 & (0.56, & 0.81) &  & 0.34 & (0.28, & 0.39) \\ 
    $a_{31}$ & 0.64 & (0.02, & 2.39) &  & 0.79 & (0.06, & 1.89) &  & 0.53 & (0.03, & 1.44) \\ 
    $a_{32}$ & -0.39 & (-3.91, & 3.56) &  & 0.71 & (-0.14, & 1.64) &  & 0.63 & (-0.07, & 1.45) \\ 
    $a_{33}$ & 2.42 & (0.15, & 6.89) &  & 0.67 & (0.03, & 1.89) &  & 0.64 & (0.03, & 1.76) \\ 
	\midrule
$\delta_2$ & 0.08 & (-0.62, & 0.77) &  & -0.37 & (-0.74, & 0.02) &  & 0.22 & (0.10, & 0.33) \\ 
  $\delta_3$ & 0.03 & (-0.76, & 0.80) &  & -0.34 & (-0.81, & 0.15) &  & 0.10 & (-0.52, & 0.72) \\
	\midrule 
  $\beta_{1,\mathrm{tod=n}}$ & -0.40 & (-0.46, & -0.35) &  & -0.20 & (-0.25, & -0.15) &  & -0.18 & (-0.23, & -0.13) \\
  \smallskip
  $\beta_{1,\mathrm{tod=m}}$ & 0.12 & (-0.00 & 0.25) &  & -0.12 & (-0.26, & 0.02) &  & 0.15 & (0.01, & 0.29) \\
  $\beta_{2,\mathrm{tod=n}}$ & -0.02 & (-0.11, & 0.06) &  & -0.07 & (-0.17, & 0.02) &  & -0.12 & (-0.21, & -0.03) \\
  $\beta_{2,\mathrm{tod=m}}$ & 0.01 & (-0.07, & 0.09) &  & -0.04 & (-0.13, & 0.06) &  & 0.01 & (-0.08, & 0.11) \\ 
  $\beta_{2,\mathrm{dow=e}}$ & -0.06 & (-0.14, & 0.02) &  & -0.02 & (-0.11, & 0.08) &  & -0.23 & (-0.32, & -0.14) \\ 
  \smallskip
  $\beta_{2,\mathrm{dow=m}}$ & -0.04 & (-0.11, & 0.04) &  & -0.01 & (-0.10, & 0.07) &  & -0.11 & (-0.19, & -0.03) \\ 
  $\beta_{3,\mathrm{tod=n/m}}$ & -0.17 & (-0.75, & 0.42) &  & -0.03 & (-0.73, & 0.70) &  & 0.04 & (-0.59, & 0.66) \\ 
  $\beta_{3,\mathrm{dow=e/m}}$ & -0.07 & (-0.64, & 0.50) &  & -0.15 & (-0.80, & 0.50) &  & -0.34 & (-0.94, & 0.25) \\
	\midrule
  $\gamma_1$ & -1.84 & (-1.94, & -1.75) &  & -1.98 & (-2.07, & -1.88) &  & -1.93 & (-2.00, & -1.87) \\ 
  $\gamma_2$ & -1.25 & (-1.33, & -1.17) &  & -1.33 & (-1.40, & -1.27) &  & -1.26 & (-1.32, & -1.20) \\ 
  $\gamma_3$ & -0.50 & (-0.57, & -0.43) &  & -0.42 & (-0.48, & -0.37) &  & -0.25 & (-0.31, & -0.20) \\ 
  $\gamma_4$ & 0.72 & (0.64, & 0.80) &  & 0.72 & (0.66, & 0.79) &  & 0.65 & (0.60, & 0.70) \\ 
	 \bottomrule[1.5pt]
\end{tabular}
\end{table}
The regression coefficients for the two models are comparable. The baseline category is residential locations during the daytime. 
For all three components, we estimate a lower level in the latent process (i.e. higher probability of a lower rating) for residential locations at night. We are able to estimate this effect as caregivers were asked to rate their home neighborhood on all three components of collective efficacy for both the daytime and the nighttime. For nonresidential and ``other'' locations, we do not explicitly have this counter factual information in the data set. For the land-use mean shift parameters in the land-use filtering model, we estimate an increase in observation for nonresidential locations compared to
residential locations. Additionally, there is some evidence of an negative effect on trust for nonresidential locations compared to residential locations. The stationary covariance model estimates more clear effects (credible intervals that do not contain zero) of the mean shift parameters which may arise due to over-smoothing across land-use categories.

\begin{table}[ht]
\caption{Estimated posterior means and corresponding posterior 95\% credible intervals of the parameters in the stationary Bayesian ordinal regression model. The numeric subscripts for the $\beta$'s and $\delta$'s correspond to the land-use processes: 1 for residential, 2 for nonresidential, 3 for the ``other'' category. The subscripts ``tod'' and ``dow'' stand for time of day and day of week with ``n'' for \textbf{n}ight, ``m'' for \textbf{m}ixed, and ``e'' for week\textbf{e}nd.}
\label{tab:stationary}
 \setlength\tabcolsep{4pt}
\centering
\begin{tabular}{lrr@{\hskip 1pt}rrrr@{\hskip 1pt}rrrr@{\hskip 1pt}r}
	\toprule[1.5pt]
	& \multicolumn{3}{c}{Defense} &	& \multicolumn{3}{c}{Trust} & & \multicolumn{3}{c}{Observation} \\
	\cmidrule{2-4} \cmidrule{6-8} \cmidrule{10-12}
	& \shortstack[r]{Posterior \\ mean} & \multicolumn{2}{c}{\shortstack[r]{95\% Credible \\ interval}} 
	& & \shortstack{Posterior \\ mean} & \multicolumn{2}{c}{\shortstack{95\% Credible \\ interval}} 
	& & \shortstack{Posterior \\ mean} & \multicolumn{2}{c}{\shortstack{95\% Credible \\ interval}} \\
  \midrule
$\phi$ & 103.21 & (69.8, & 140.0) &  & 120.15 & (86.5, & 155.1) &  & 494.95 & (348.4, & 716.5) \\ 
  $\tau^2$ & 0.33 & (0.26, & 0.42) &  & 0.79 & (0.66, & 0.92) &  & 0.31 & (0.25, & 0.38) \\
	\midrule
$\delta_2$ & 0.09 & (0.01, & 0.16) &  & -0.33 & (-0.41, & -0.24) &  & 0.16 & (0.09, & 0.23) \\ 
  $\delta_3$ & 0.09 & (-0.24, & 0.41) &  & -0.22 & (-0.58, & 0.15) &  & 0.09 & (-0.54, & 0.71) \\ 
	\midrule 
$\beta_{1,\mathrm{tod=n}}$ & -0.41 & (-0.48, & -0.34) &  & -0.16 & (-0.22, & -0.09) &  & -0.24 & (-0.30, & -0.18) \\ 
\smallskip
  $\beta_{1,\mathrm{tod=m}}$ & 0.13 & (-0.01, & 0.26) &  & -0.09 & (-0.23, & 0.06) &  & 0.09 & (-0.05, & 0.23) \\
  $\beta_{2,\mathrm{tod=n}}$ & -0.04 & (-0.12, & 0.05) &  & -0.08 & (-0.17, & 0.01) &  & -0.12 & (-0.21, & -0.03) \\
  $\beta_{2,\mathrm{tod=m}}$ & 0.01 & (-0.08, & 0.09) &  & -0.03 & (-0.12, & 0.06) &  & 0.01 & (-0.08, & 0.10) \\ 
  $\beta_{2,\mathrm{dow=e}}$ & -0.06 & (-0.14, & 0.03) &  & -0.02 & (-0.12, & 0.07) &  & -0.23 & (-0.32, & -0.14) \\ 
  \smallskip
  $\beta_{2,\mathrm{dow=m}}$ & -0.00 & (-0.08, & 0.07) &  & 0.00 & (-0.08, & 0.09) &  & -0.10 & (-0.18, & -0.02) \\ 
  $\beta_{3,\mathrm{tod=n/m}}$ & -0.10 & (-0.76, & 0.55) &  & -0.13 & (-0.83, & 0.59) &  & 0.02 & (-0.62, & 0.66) \\ 
  $\beta_{3,\mathrm{dow=e/m}}$ & 0.03 & (-0.57, & 0.64) &  & -0.17 & (-0.80, & 0.45) &  & -0.40 & (-1.00, & 0.20) \\ 
	\midrule
  $\gamma_1$ & -1.97 & (-2.02, & -1.92) &  & -1.77 & (-1.86, & -1.67) &  & -2.00 & (-2.04, & -1.95) \\ 
  $\gamma_2$ & -1.34 & (-1.37, & -1.30) &  & -1.13 & (-1.20, & -1.06) &  & -1.34 & (-1.37, & -1.30) \\ 
  $\gamma_3$ & -0.55 & (-0.57, & -0.53) &  & -0.24 & (-0.31, & -0.16) &  & -0.30 & (-0.33, & -0.29) \\ 
  $\gamma_4$ & 0.75 & (0.72, & 0.78) &  & 0.90 & (0.84, & 0.95) &  & 0.59 & (0.57, & 0.61) \\ 
	 \bottomrule[1.5pt]
\end{tabular}
\end{table}

\subsection{Predictions}
\begin{figure}[ht]
\begin{minipage}{.5\textwidth}
	\centering
	\includegraphics[width=.85\linewidth]{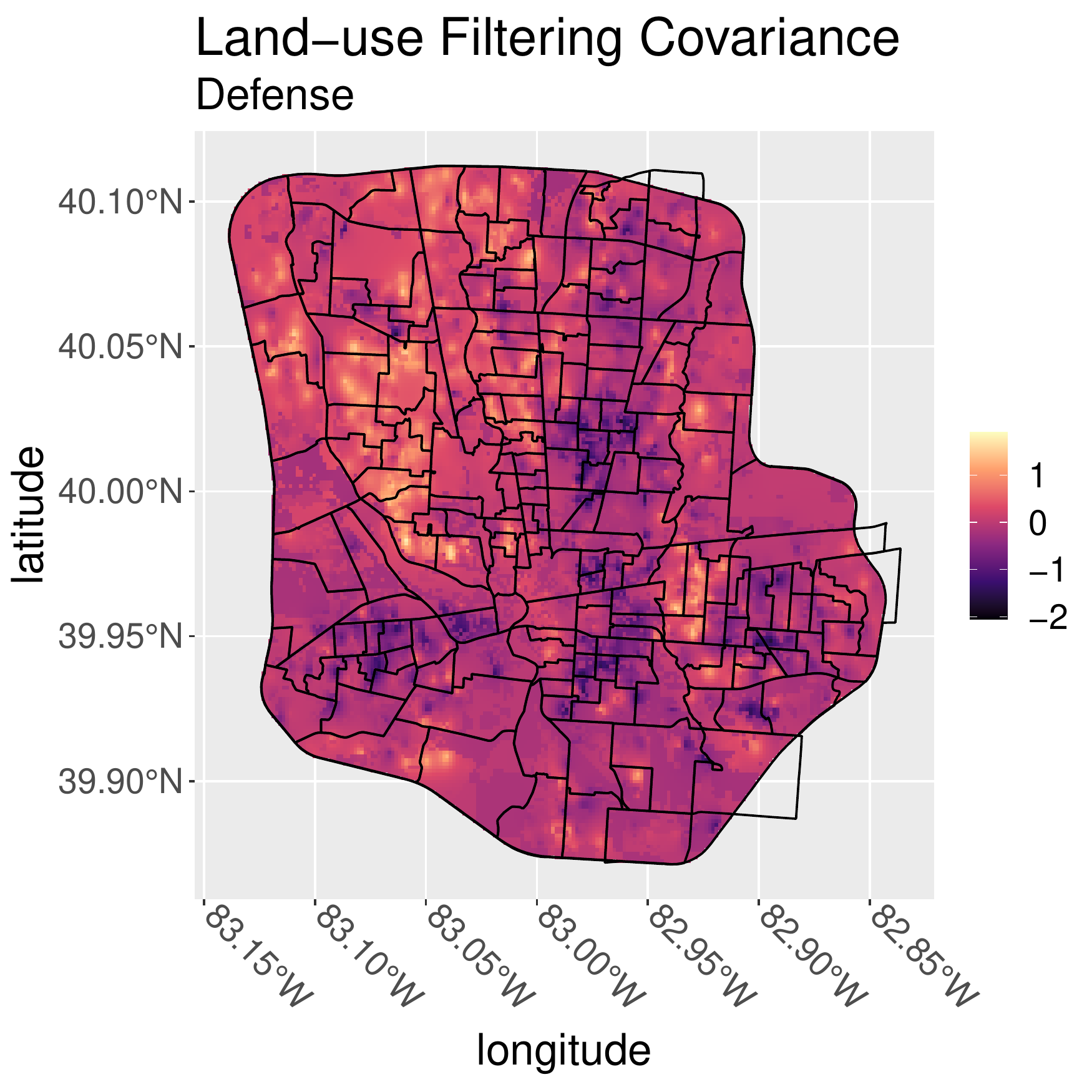}
\end{minipage}%
\begin{minipage}{.5\textwidth}
	\centering
	\includegraphics[width=.85\linewidth]{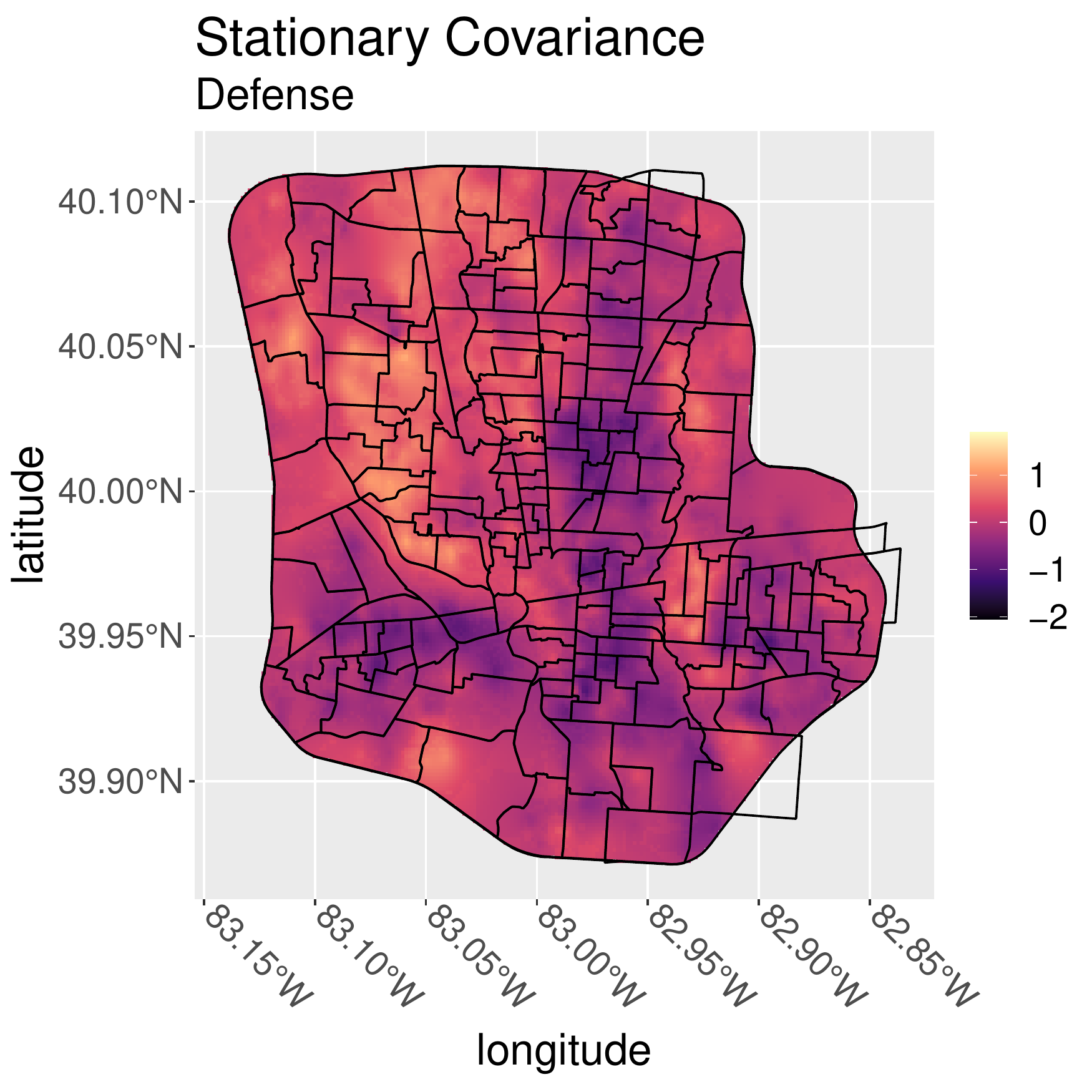}
\end{minipage}
\begin{minipage}{.5\textwidth}
	\centering
	\includegraphics[width=.85\linewidth]{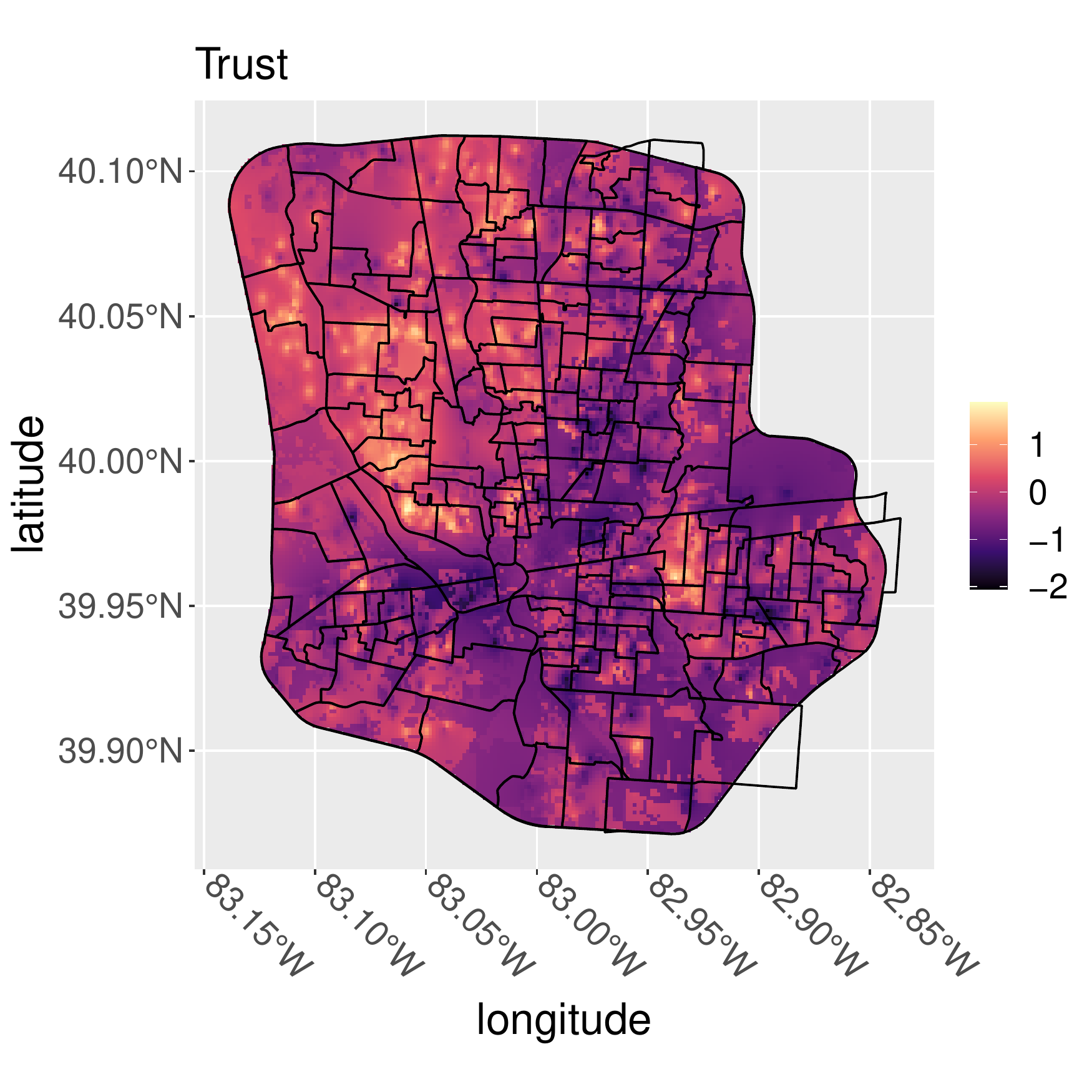}
\end{minipage}%
\begin{minipage}{.5\textwidth}
	\centering
	\includegraphics[width=.85\linewidth]{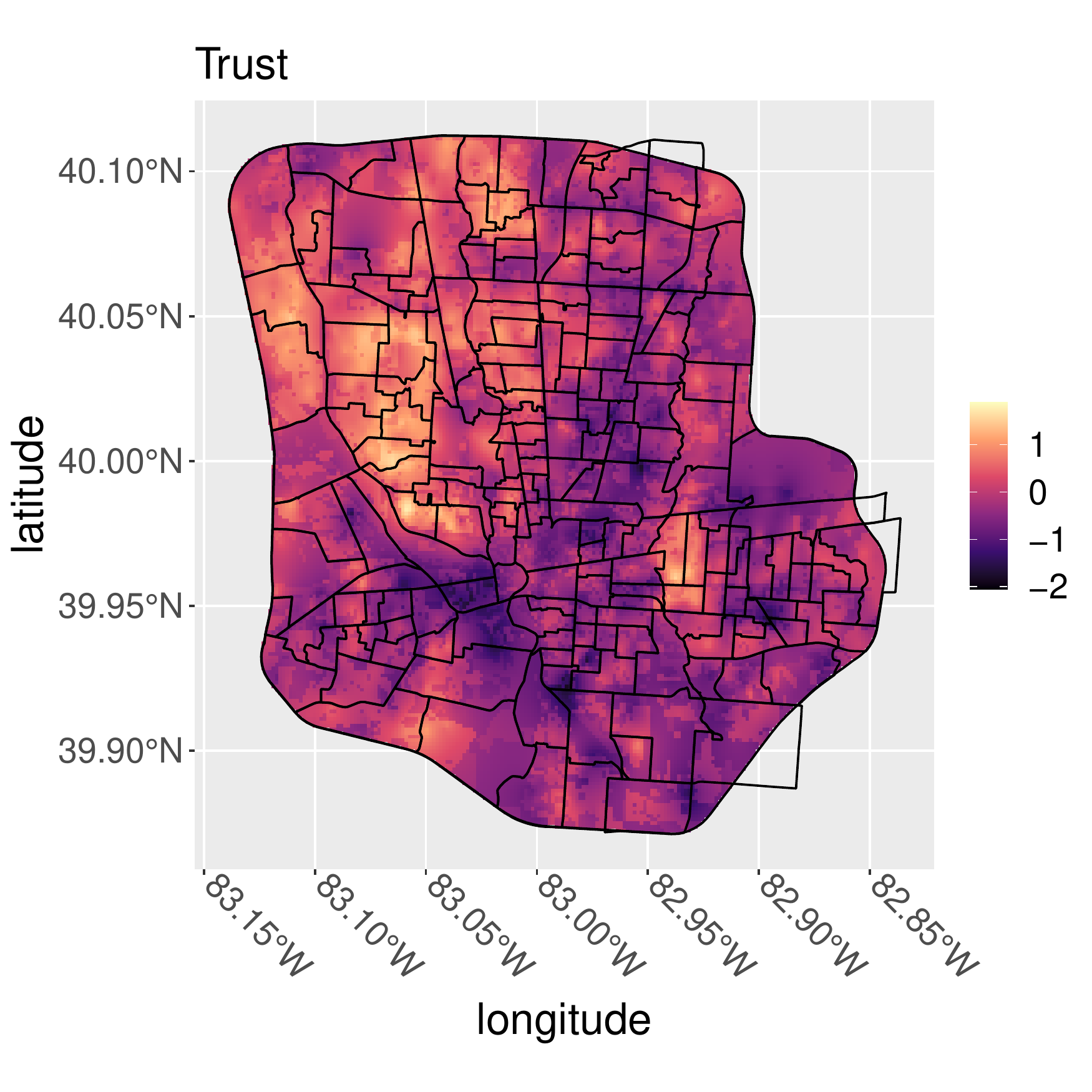}
\end{minipage}
\begin{minipage}{.5\textwidth}
	\centering
	\includegraphics[width=.85\linewidth]{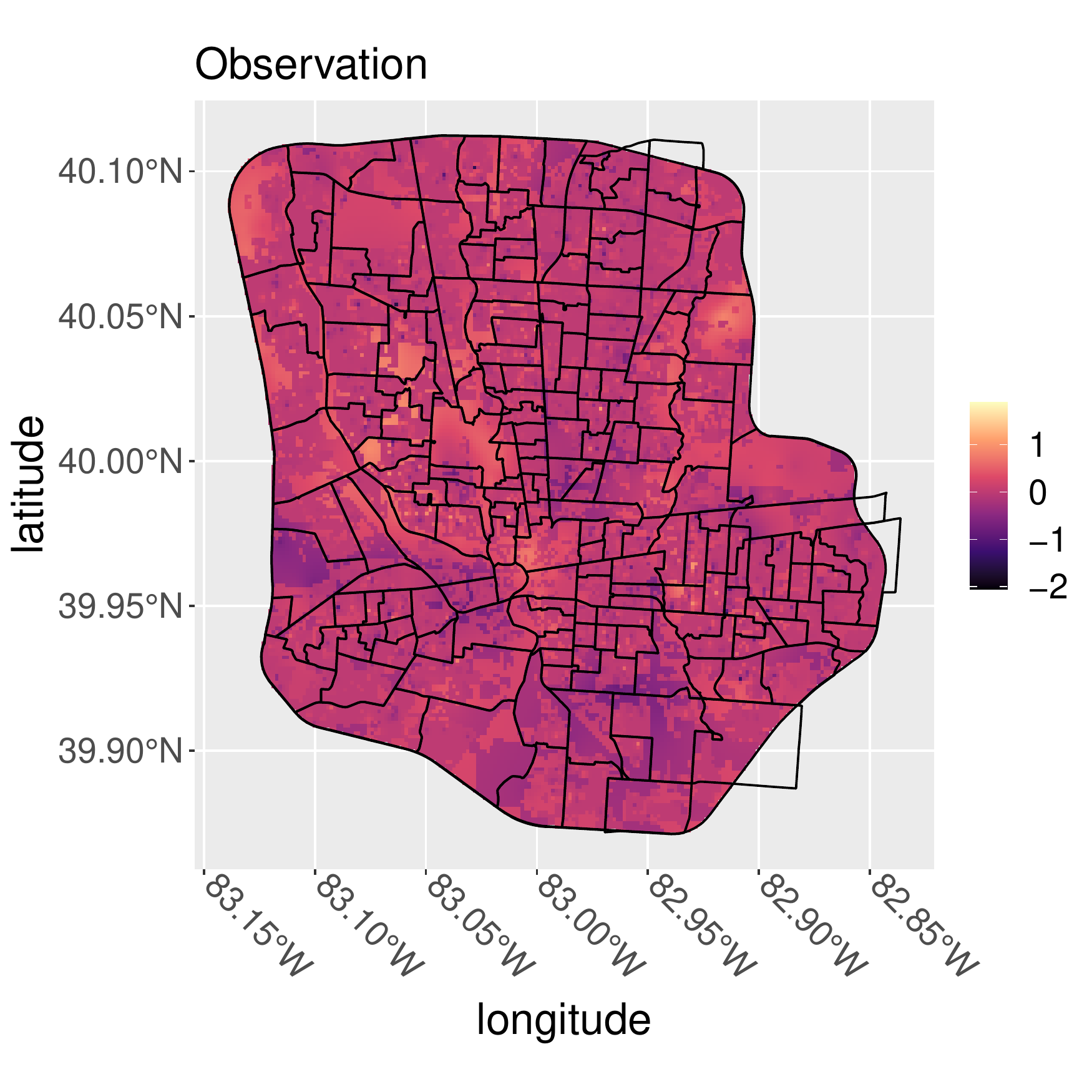}
\end{minipage}%
\begin{minipage}{.5\textwidth}
\centering
	\includegraphics[width=.85\linewidth]{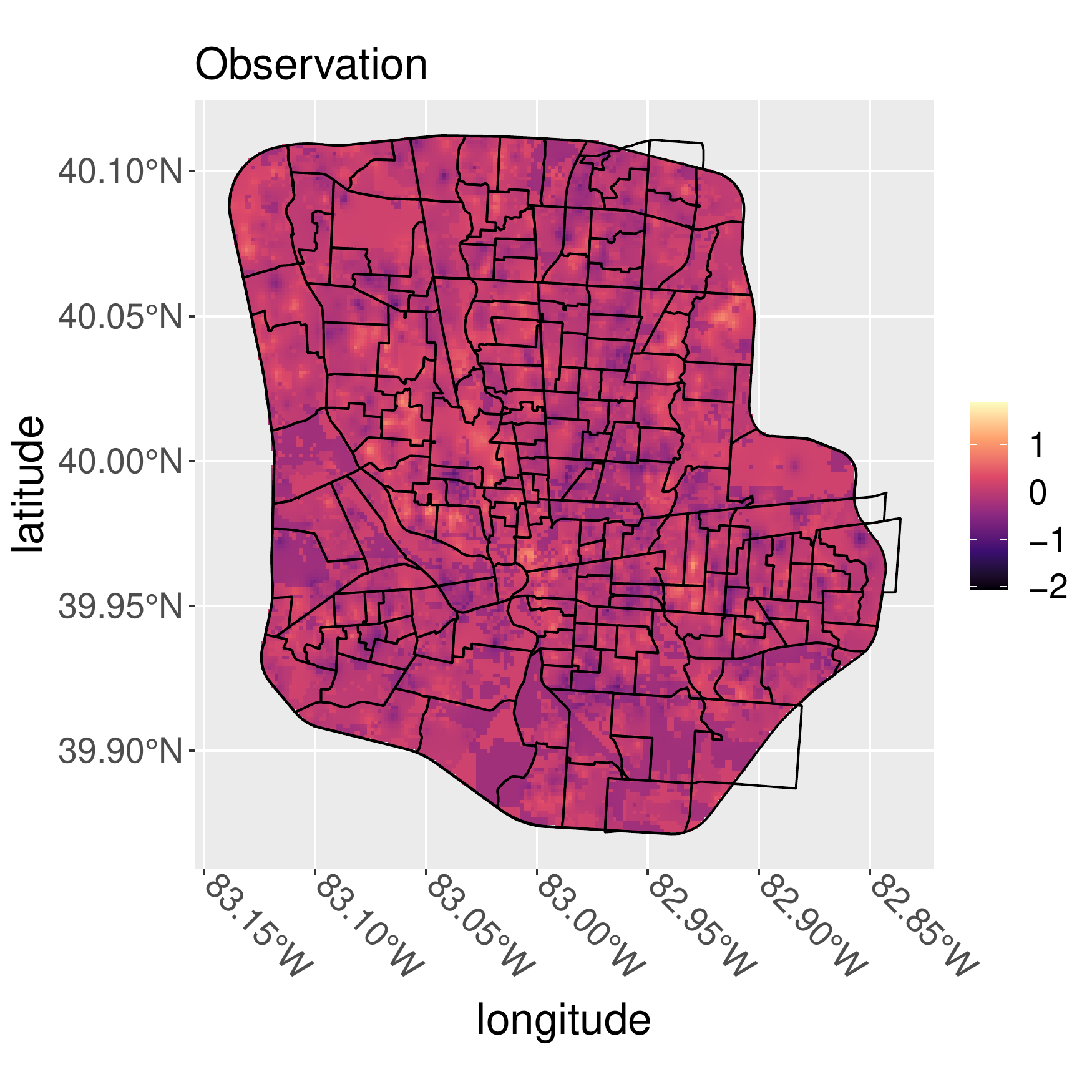}
%
%\end{center}
\end{minipage}
\caption{Point-wise posterior means of the spatial random effects for defense (top row), trust (middle row), and observation (bottom row). On the left are the predictions of the collective efficacy components derived from the land-use filtering model.  Predictions from the stationary model are on the right.  The scales are the same for plots on the same row.}
\label{fig:pred_all}
\end{figure}

Visualization of the predicted levels of each component of collective efficacy also supports the use of the land-use filtering model. Figure~\ref{fig:pred_all} contains the point-wise posterior mean predictions of the spatial random effect $\tilde\Z$ across a fine grid of points (around 30,000) within the I-270 belt loop. The land-use category of each prediction location was assigned by the land-use category of the nearest parcel.  We obtained predictions using Equation~\ref{eq:post_spat_re} for each of the draws of the thinned MCMC chains used in the WAIC calculations. Plotted is the point-wise posterior mean of the spatial random effect ($\mubf^{[b]}$ in Equation~\ref{eq:post_spat_re}) at each prediction location averaged across the thinned draws. We calculated the posterior mean of the spatial random effect rather than the mean of posterior predictive distribution due to memory constraints for obtaining a multivariate normal sample at a very fine grid of locations. The fine grid of locations allows us to better visualize within neighborhood variation across land-use categories. The predicted stationary process on the right is more smooth compared to the predicted land-use process on the left. Land-use filtering allows use to estimate distinct processes for each land-use type while also smoothing over land-use boundaries. There are discontinuities in the prediction at the boundaries from one land-use type to another, however, these discontinuities would be greater if we simply fit three independent process (i.e. the matrix of weights $\A$ is a diagonal matrix) for each land-use category and then filtered the processes. The predicted land-use filtered process for the defense and trust components shows the shorter range spatial dependence within residential areas and the longer range spatial dependence for non-residential locations. For the observation component, the spatial range of residential locations is effectively zero, thus the predictions revert to the mean for residential locations while allowing for greater spatial smoothing over nonresidential locations. 

As the full Bayesian spatial ordinal regression model is time consuming to fit we also compared the predictions from the ordinal land-use filtering model to predictions from the approximate land-use filtering model described in Section~\ref{sec:amap}. The approximate model treats the rating response directly as a Gaussian outcome, thus the scales of predictions are different. 
Additionally, treating the outcome directly as Gaussian allows one to drop the restrictions on the covariance function when defining the land-use filtering process. Predictions from both models are very highly correlated for each component of collective efficacy: predictions for the defense component had a correlation of 0.93, for trust a correlation of 0.90, and for observation a correlation of 0.92. In practice, when uncertainty estimates of the collective efficacy process are not needed, using the approximate model will give nearly the same results as the full Bayesian spatial ordinal model with a significant reduction in computation time. 

\section{Simulation Study}
\label{sec:sim_study}
To further understand the uses and limitations of our model we explore model 
performance when data are generated from a land-use filtering process as described 
in Section~\ref{sec:methods}. In this model generated data setting we seek to understand the limits of our partially identified model. 

As we have a partially identified model, the constraints explained in Section~\ref{sec:ident} do not guarantee that we will be able to recover the values of model parameters used to generate the data. Indeed, all spatial probit regression models are partially
identifiable: a direct result of binning the latent spatial process 
into ordinal or binary categories. While this loss of information weakens identifiability of the spatial parameters, we can still evaluate the predictive performance of our model by comparing 
predictions from the model to the true latent process of the generative model over a 
fine grid of points. This evaluation of model performance aligns with the main goal 
of our analysis to understand the fine grain variation of collective efficacy within 
census units. 

For our simulation study we use the same parameter settings as in the demonstration
of the generative model in section~\ref{sec:motivation}. Our test set to evaluate 
predictive performance is the same fine grid of ten thousand points
over the unit square domain. For each iteration of the simulation study we randomly
generate a training set of one thousand $x$ and $y$ coordinates 
drawn uniformly from the unit square. Each training set location is assigned the 
land use category of the nearest test set location. Next, we draw a realization of the latent variable $\mathbf{Z}^*$ at the combined set of test and training locations with mean zero 
and covariance function defined by equation~\ref{eq:vec_sglmm} with 
$$
\A = 
\begin{bmatrix} 1.8 & 0.0 & 0.0 \\
0.8 & 1.2 & 0.0 \\
0.9 & 1.0 & 1.25 
\end{bmatrix},\;\;
\phibf =\begin{bmatrix}40.0 & 10.0 & 2.0\end{bmatrix}',\text{ and }
\sigmabf^2 =\begin{bmatrix}1.0 & 1.0 & 1.0\end{bmatrix}'.
$$
 From the latent process realization we apply the cutoffs $\gammabf
= \begin{bmatrix}
-2.0 & -0.5 & 0.1 & 1.1
\end{bmatrix}' $ to obtain the ordinal response. Then we estimate the parameters of the 
approximate model as described in section~\ref{sec:amap} for both the land-use 
filtering and stationary specification (note that for both models $\deltabf$ and 
$\betabf$ are set to be zero). Using the approximate maximum a posteriori estimates 
we the generate predictions for the test set conditional on the observed ordinal 
responses. We use the approximate model for computational efficiency and because the approximate model predictions are highly correlated with the full Bayesian model predictions. Since the predictions conditional on the ordinal data will be on a 
different scale than the latent process, to evaluate model performance we compare 
the centered and scaled predictions given the ordinal response to the true latent 
process values at the test set of locations. Out of 100 replications of the simulation study (generating data from the land-use filtering model and fitting both the approximate land-use model and stationary model to the simulated data at the training locations)  the land-use filtering model had a lower mean (over the test set of locations) 
absolute error in 76\%  of the replications and higher 
correlation coefficient between the predictions and true values at the test locations in 81\% of the replications. By both the mean absolute 
error and correlation coefficient, the land-use filtering model consistently 
performs better than the stationary spatial model in predicting the components of collective efficacy at unobserved locations even though the model is only partially identifiable.

\section{Discussion} 
\label{sec:discuss}
As demonstrated in the analysis section, our proposed land-use filtering model for the AHDC collective efficacy data better fits the observed data than the default approach using a spatial generalized linear model with a single stationary latent process.  This superiority holds for all three components of collective efficacy. Due to the lack of software for fitting alternative models with a single nonstationary latent process, as opposed to the default latent stationary model discussed in Section~\ref{sec:results}, our model comparison exercise is somewhat limited in scope.  For example, \citet{risser_calder2015}'s covariance-regression model could be used to allow the parameters of the latent spatial process to vary smoothly with spatially-referenced land-use covariate information.

In our analysis, we fit our model separately to the ratings of each component of collective efficacy.  While one could consider modeling the ratings of the three components jointly
as a multivariate response, in our opinion, doing so would limit the applicability of our findings in addressing important sociological questions. Collective efficacy was originally conceived as a multidimensional construct comprised of 
separately operating social processes \citep{sampson_etal1999}. 
%Modern studies looking at antecedents and consequences of collective efficacy seek to test sociological theory involving the components individually or in pairs \citep{hipp2016, wickes_etal2019}.
Modern studies of crime and other outcomes have looked at the effects of separate components of collective efficacy \citep{hipp2016, wickes_etal2019}.
Therefore, as an input to downstream analyses, separate predictions of the component processes will allow for a richer and more nuanced collection of questions about collective efficacy to be addressed. 

As we had very few reports on locations that fell into the ``other'' category, we choose to fix $\phi_3 = \infty$ the spatial dependence parameter of the third latent process in the LMC construction. This had the effect of adding additional independent error to the spatial process for the components of collective efficacy at the ``other'' locations. Alternatively, one could consider dropping the third latent process entirely by fixing $a_{33}=0$, so that the spatial process of our three land-use categories is an expansion of a two dimensional latent spatial process. We choose to keep $\phi_3$ in the model to represent our belief that there truly are three underlying, land-use-specific processes for the components of collective efficacy, and we fix $\phi_3$ due to data availability.

% use of we do not believe doing so would be consistent each component of collective efficacy separately, as this approach is the most general form of collective efficacy modeling.
% Collective efficacy was originally conceived as a multidimensional construct with 
% separately operating social processes \citep{sampson_etal1997}. Sociologists
% are careful in evaluating how the different measures of trust, defense, and 
% observation, what we call the components of collective efficacy, relate to each other and place emphasis on distinguishing the antecedents and consequences of 
% collective efficacy \citep{hipp2016, wickes_etal2019}. 
% While one could consider modeling the 
% components of collective efficacy jointly as a multivariate response, fitting the components of 
% collective efficacy separately provides baseline 
% predictions which can then be used in other models. To expand our model one could expand the LMC formulation to jointly account for two or more of the components of collective efficacy together. This would greatly increase the number of parameters estimated in the $\A$ matrix, which in practice could be infeasible, thus hierarchical schemes within the LMC framework could be explored to reduce the dimension of the parameter space.

A limitation of our analysis is that we were not able to account for dependence in the collective efficacy component ratings made by the same individual. Conceptually, it may be possible to account for differences in how individuals use the component response scales since individuals report on several locations, which may be rated by multiple individuals. See \citet{linero_etal2018} for one approach to address individual rater effects.  However, since many routine locations will be closer to the caregiver's home location, we found that an individual random effect is essentially confounded with the spatial random effect.  Since the latter is of primary interest and necessary for spatial prediction at unrated locations, we decided to forego an individual-level random effect in our analyses.

Finally, we note that our prediction of the collective efficacy components across the study area implicitly assume that the caregivers who rate locations other than their residence are a representative sample of individuals who frequent the locations as part of their daily routine. We view this assumption as foundational to our notion of collective efficacy as a continuously-indexed spatial process.  Alternatively, one could define collective efficacy of a location from the perspective of an objective observer, who may or may not spend time at the location.  This notion is not compatible with the AHDC Study design and, more importantly, is arguably not consistent with sociological theory which emphasizes perceptions of space among users of the space.

In future work, we will use the estimated continuously-indexed spatial collective efficacy component processes to examine the relationship between collective efficacy and point-level crime data. We will also explore the hypothesis that within-neighborhood variation in collective efficacy relates to crime \citep{weisburd_etal2016}.  Such an analysis is only possible with our fine-grained maps of collective efficacy across the study area. 

In conclusion, while our proposed land-use filtering model is a novel method for learning about small-scale variability in collective efficacy across an urban environment, we acknowledge that the approach may be more generally applicable.  Other spatial prediction problems in which the spatial dependence structure may depend on land use (e.g., pollution mapping) might benefit from the modeling approach.  More broadly, any spatially-dependent, categorical variable that defines a partition of a well-defined study area could be used instead of land use in the approach.  

\begin{funding}
This study was supported in part by the National Institute on Drug Abuse (Christopher R. Browning; R01DA032371); the Eunice Kennedy Shriver National Institute on Child Health and Human Development (Catherine A. Calder; R01HD088545; John Casterline, The Ohio State University Institute for Population Research, P2CHD058484; Elizabeth Gershoff, The University of Texas at Austin Population Research Center, P2CHD-042849); and the W. T. Grant Foundation.
\end{funding}

%%%%%%%%%%%%%%%%%%%%%%%%%%%%%%%%%%%%%%%%%%%%%%
%% Supplementary Material, including data   %%
%% sets and code, should be provided in     %%
%% {supplement} environment with title      %%
%% and short description. It cannot be      %%
%% available exclusively as external link.  %%
%% All Supplementary Material must be       %%
%% available to the reader on Project       %%
%% Euclid with the published article.       %%
%%%%%%%%%%%%%%%%%%%%%%%%%%%%%%%%%%%%%%%%%%%%%%
\begin{supplement}
\stitle{Example dataset and code} \sdescription{AHDC data will be deposited to Inter-university Consortium for Political and Social Research (ICPSR) in publicly available and restricted access forms (expected in Summer 2023). To ensure participant privacy and maintenance of data confidentiality, the ADHC caregiver location reports needed to reproduce the analyses in this paper will only be available in the restricted access version of the data. Qualified researchers will be able to submit an application to ICPSR for access to the data.  In our supplementary material, we instead provide a synthetic data set generated from the posterior predictive distribution of the trust component at randomly selected locations out of the grid of points used for our predictive maps. Accompanying the example data set is code to reproduce the analyses done in the paper.}
\end{supplement}

%%  imsart-nameyear.bst  will be used to                   %%
%%  MR numbers will be added by VTeX.                      %%
%%                                                         %%
%%  Use \cite{...} to cite references in text.             %%
%%                                                         %%
%%%%%%%%%%%%%%%%%%%%%%%%%%%%%%%%%%%%%%%%%%%%%%%%%%%%%%%%%%%%%

%% if your bibliography is in bibtex format, uncomment commands:
\clearpage
\bibliographystyle{imsart-nameyear} % Style BST file
\bibliography{references.bib}       % Bibliography file (usually '*.bib')

%% or include bibliography directly:
% \begin{thebibliography}{}
% \bibitem[\protect\citeauthoryear{???}{???}]{b1}
% \end{thebibliography}

\end{document}